 \documentclass{aa}
\usepackage[dvips]{graphicx}
\usepackage{psfig}
\begin{document}
\title{New neighbours: V. 35 DENIS late-M dwarfs \\
          between 10 and 30 parsecs}


   \author{N.~Phan-Bao
       \inst{1,2}
       \and
       F.~Crifo
       \inst{2}
       \and
       X.~Delfosse
       \inst{3}
       \and
       T.~Forveille
       \inst{3,4}
       \and
       J.~Guibert
       \inst{1,2}
       \and
       J.~Borsenberger
       \inst{5}
       \and
       N.~Epchtein
       \inst{6}
       \and
       P.~Fouqu\'e
       \inst{7,8}
       \and
       G.~Simon
       \inst{2}
       \and
       J.~Vetois
       \inst{1,9}
       }

   \offprints{F. Crifo, \email{francoise.crifo@obspm.fr}}

   \institute{Centre d'Analyse des Images, GEPI, Observatoire de Paris,  
                61 avenue de l'Observatoire, 75014 Paris, France
             \and
              GEPI, Observatoire de Paris, 5 place J. Janssen, 92195 
              Meudon Cedex, France
             \and
              Laboratoire d'Astrophysique de Grenoble, Universit\'e J. 
              Fourier, B.P. 53, F-38041 Grenoble, France
             \and
              Canada-France-Hawaii Telescope Corporation, 65-1238 Mamalahoa 
              Highway, Kamuela, HI 96743 USA
             \and
              SIO, Observatoire de Paris, 5 place J. Janssen, 92195 
              Meudon Cedex, France
             \and
              Observatoire de la Cote d'Azur, D\'epartement Fresnel, BP 4229, 
                06304 Nice Cedex 4, France
             \and
              LESIA, Observatoire de Paris, 5 place J. Janssen, 92195 
              Meudon Cedex, France
             \and
              European Southern Observatory, Casilla 19001, Santiago 19, Chile
             \and
              Ecole Normale Sup\'erieure de Cachan, 61 avenue du 
              Pr\'esident-Wilson, 94230 
                Cachan, France
             }

   \date{Received / Accepted}

\abstract{
This paper reports updated results on our systematic mining of the DENIS 
database for nearby very cool M-dwarfs (M6V-M8V, $2.0 \leq I-J \leq 3.0$, 
photometric distance within 30~pc), initiated by Phan-Bao et al. 
(\cite{phan-bao}, hereafter Paper I). We use M dwarfs with well measured 
parallaxes (HIP, GCTP,...) to calibrate the DENIS ($M_{\rm I}$, $I-J$) 
colour-luminosity relationship. The resulting distance error for single 
dwarfs is about 25\%. 
Proper motions, as well as B and R magnitudes, were measured on archive 
Schmidt plates for those stars in the DENIS database that meet the 
photometric selection criteria.
We then eliminate the giants by a Reduced Proper Motion cutoff, which is
significantly more selective than a simple proper motion cutoff. It
greatly reduces the selection bias against low tangential velocity
stars, and results in a nearly complete sample. \\
     Here we present new data for 62 red dwarf candidates selected 
over 5700 square degrees in the DENIS database. 26 of those originate
in the 2100 square degrees analysed in Paper~I, with improved parameters 
here, and 36 were found in 3600 additional square degrees. 25 of those 
are new nearby dwarfs.
We determine from that sample of 62 stars a stellar density for 
$12.0 \leq M_{\rm I} \leq 14.0$ of 
$\overline{\Phi}_{\rm I~cor}=(2.2 \pm 0.4).10^{-3}$stars.pc$^{-3}$.mag$^{-1}$.
This value is consistent with photometric luminosity functions measured
from deeper and smaller-field observations, but not with the nearby star
luminosity function. \\
In addition we cross-identified the NLTT and DENIS catalogues to 
find 15 similar stars, in parts of the sky not yet covered by the 
colour-selected search. 
We present distance and luminosity estimates for these 15
stars,  10 of which are newly recognized nearby dwarfs. 
A similar search in Paper~I produced 4 red dwarf candidates, and
we have thus up to now identified a total of 35 new nearby late-M dwarfs.
\keywords{Astrometry - proper motions - very low mass stars - solar neighbourhood}
}

\titlerunning{New neighbours: V. 35 DENIS late-M dwarfs}
\authorrunning{N. Phan-Bao et al.}
\maketitle


\section{Introduction}

The stellar content of the solar neighbourhood is once again a very active 
research field, revived in large part by the vast amounts of new data from 
the near-Infrared surveys DENIS (Epchtein \cite{epchtein97}) and 2MASS
(Skrutskie et al. \cite{skru97}) and the optical Sloan Digital Sky Survey 
(York et al. \cite{york00}, Hawley et al. \cite{hawley}).
These surveys have identified much fainter and cooler objects, and
required the extension of the spectral classification system by two
new spectral classes, the L and T dwarfs (Mart\'{\i}n et al. \cite{martin97}; 
Kirkpatrick et al. \cite{kirkpatrick99}). As expected, the surveys also
detect large numbers of less extreme late-M dwarfs. As shown by Gliese 
et al. (\cite{gliese86}) the census of the solar neighbourhood is rather
incomplete for late M dwarfs, and their actual number density is not very 
well established.

In Paper I (Phan-Bao et al. \cite{phan-bao}), we 
presented 30 nearby ($d_{\rm phot} < 30$~pc) late-M dwarfs 
($2.0 \leq I-J \leq 3.0$, M6-M8) 
with high proper motions: 26,
a few of which were previously known from other sources, were photometrically
selected from 2100 square degrees of DENIS data, and 4 were identified by 
cross-identifying the LHS (Luyten \cite{luytena}) and DENIS catalogues
over a larger sky area. 
Here we repeat the analysis of Paper~I with an improved ($I-J$, $M_{\rm I}$)
relation, calibrated specifically for the DENIS filter set, and extend 
the colour selection to a further 3600 square degrees.
We also use an improved dwarf/giant discrimination criterion, based on the 
reduced proper motion rather than the simple 
proper motion cutoff which is commonly used for that purpose (e.g. Scholz et 
al. \cite{scholz}, and Paper~I). This allows us to dig down to significantly 
lower proper motions, and thus to identify additional dwarf candidates. 
Finally, we systematically search the DENIS database for southern NLTT 
stars (Luyten \cite{luytenb}) that have colours in the same 
($2 \leq I-J \leq 3.0$) range. 

Sect.~2 presents the DENIS colour-magnitude relation, and Sect.~3 reviews 
the sample selection. Section 4 discusses the proper motion 
measurements and the calibration of the  B and R photographic photometry.
Sect.~5 presents the giant/dwarf discrimination from Reduced Proper Motion
plots, and Sect.~7 a rough estimation of effective temperatures.
We discuss the completeness of the sample in Sect.~6 and indicate future
directions in Sect.~8.
%
\begin{figure}
\psfig{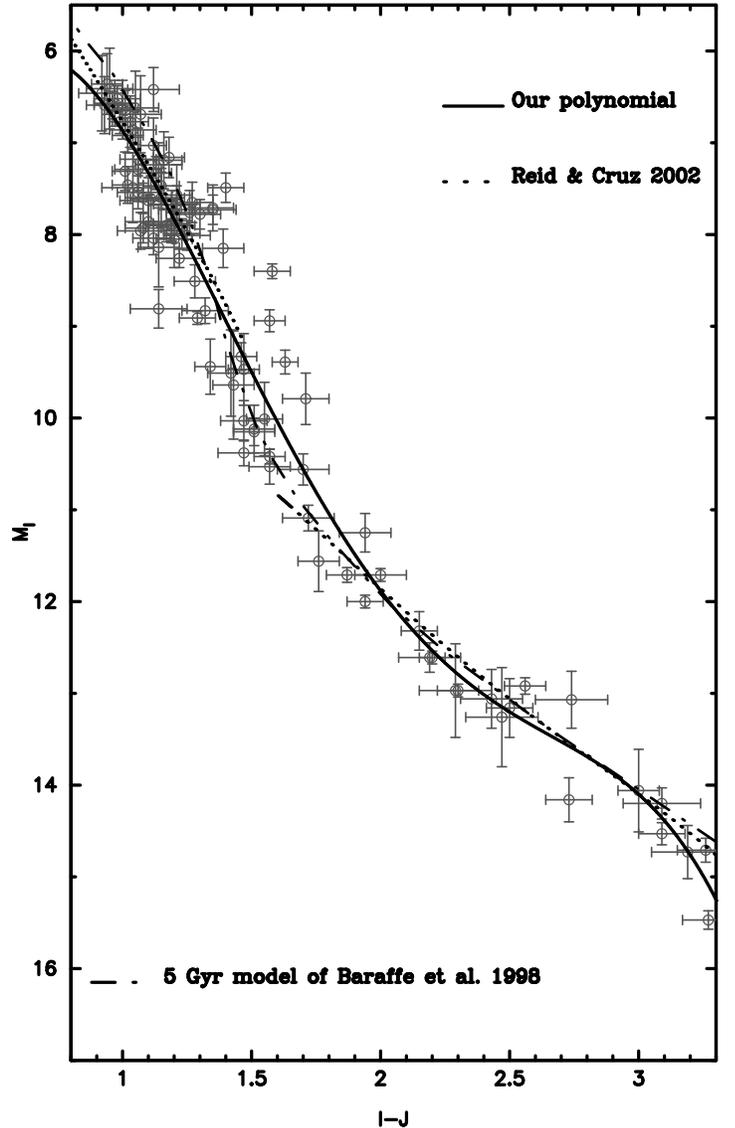} 
\caption{($M_{\rm I}, I-J$) HR diagram for single M dwarfs with known 
trigonometric parallaxes (data in Table~\ref{table_cali}).}
\label{fig_col_mag}
\end{figure}
\begin{figure}
\psfig{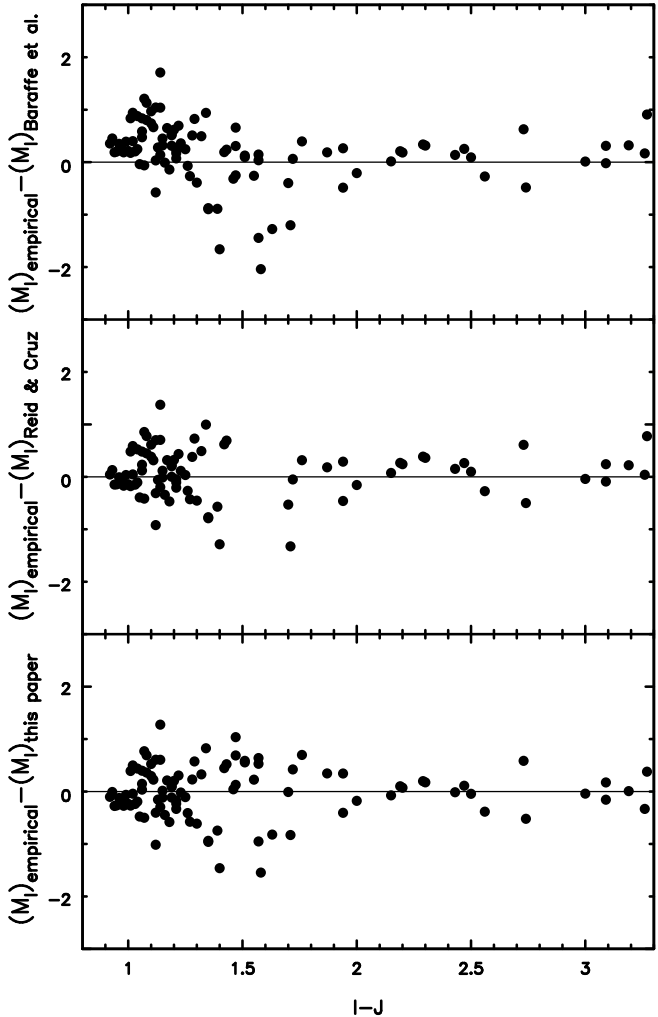}
\caption
{Empirical $M_{\rm I}$ absolute magnitudes compared with values
obtained from the theoretical tracks of Baraffe et al. (\cite{baraffe}; top), 
from the piecewise polynomial calibration of Reid \& Cruz (\cite{reid}; middle)
(except for $1.45 < I-J < 1.65$), and from the calibration derived
in this paper (bottom).
}
\label{residu}
\end{figure}

\section{DENIS colour-magnitude relation}
\begin{table*}
   \caption{Single red dwarfs with accurate trigonometric parallaxes 
    and good DENIS photometry, used for the absolute 
    magnitude calibration}
    \label{table_cali}
  $$
   \begin{tabular}{llllllllllllll}
   \hline 
   \hline
   \noalign{\smallskip}
Stars &$\alpha_{\rm 2000}$&$\delta_{\rm 2000}$&DENIS& I  &I$-$J&J$-$K&errI&errJ&errK&$\pi$&err$\pi$&$M_{\rm I}$& Ref \\ 
      &                   &                   &Epoch&    &      &    &    &    &    & mas & mas&          &     \\ 
  (1)&(2) &(3) &(4)  &(5) &(6) &(7) &(8)  &(9) &(10) &(11) &(12) &(13) &(14)\\
   \hline
   \noalign{\smallskip}  
G 158-027       & 00 06 43.37 & $-$07 32 10.5 & 1996.693 & 10.36  & 1.94 & 0.96 & 0.03 & 0.06 & 0.07 & 213.0  &~\,3.6 & 12.00  & b  \\
LHS 1026        & 00 09 04.32 & $-$27 07 19.8 & 1999.833 &~\,9.77 & 1.08 & 0.65 & 0.03 & 0.06 & 0.07 &~\,42.8 &~\,2.6 &~\,7.93 & a  \\ 
HIP 1399        & 00 17 30.41 & $-$59 57 04.3 & 1998.600 &~\,9.66 & 1.12 & 0.97 & 0.02 & 0.10 & 0.09 &~\,22.5 &~\,2.3 &~\,6.42 & a  \\ 
GJ 2003         & 00 20 08.37 & $-$17 03 40.7 & 2000.589 &~\,9.70 & 1.21 & 0.77 & 0.03 & 0.08 & 0.10 &~\,39.3 &~\,3.1 &~\,7.67 & a  \\ 
GJ 1009         & 00 21 56.03 & $-$31 24 21.9 & 1999.756 &~\,9.03 & 1.21 & 1.11 & 0.03 & 0.06 & 0.08 &~\,54.9 &~\,2.2 &~\,7.73 & a  \\ 
LHS 1064        & 00 23 18.55 & $-$50 53 38.1 & 2000.860 &~\,9.52 & 1.19 & 0.89 & 0.04 & 0.08 & 0.09 &~\,47.4 &~\,2.8 &~\,7.90 & a  \\ 
BRI 0021-02*    & 00 24 24.63 & $-$01 58 20.0 & 1998.734 & 15.13  & 3.26 & 1.31 & 0.04 & 0.10 & 0.09 &~\,82.5 &~\,3.4 & 14.71  & b  \\
LHS 1106        & 00 35 59.98 & $-$09 30 56.0 & 2000.762 &~\,9.59 & 1.04 & 0.83 & 0.03 & 0.08 & 0.08 &~\,28.5 &~\,2.3 &~\,6.86 & a  \\ 
LHS 1122        & 00 39 58.88 & $-$44 15 11.8 & 2000.688 &~\,9.48 & 1.19 & 0.83 & 0.03 & 0.09 & 0.07 &~\,43.6 &~\,2.6 &~\,7.68 & a  \\ 
LP 646-17       & 00 48 13.33 & $-$05 08 07.4 & 2000.548 &~\,9.91 & 1.10 & 0.87 & 0.03 & 0.08 & 0.08 &~\,38.9 &~\,4.7 &~\,7.86 & a  \\ 
RGO 0050-2722   & 00 52 54.67 & $-$27 05 59.5 & 1998.729 & 16.67  & 3.19 & 1.06 & 0.08 & 0.12 & 0.14 &~\,41.0 &~\,4.0 & 14.73  & b  \\ 
G 70-22         & 00 56 30.25 & $-$04 25 15.0 & 2000.603 & 12.18  & 1.71 & 0.83 & 0.03 & 0.08 & 0.06 &~\,33.3 &~\,3.8 &~\,9.79 & b  \\ 
LP 706-69       & 00 56 50.41 & $-$11 35 19.7 & 2000.603 &~\,9.63 & 1.35 & 0.89 & 0.03 & 0.08 & 0.06 &~\,41.7 &~\,2.5 &~\,7.73 & a  \\ 
G 268-110       & 01 04 53.70 & $-$18 07 29.2 & 1998.718 & 11.09  & 1.72 & 0.97 & 0.03 & 0.10 & 0.09 &~\,99.8 &~\,5.0 & 11.09  & b  \\
LP 293-94       & 01 17 59.36 & $-$48 09 01.0 & 1999.805 &~\,9.74 & 1.07 & 0.84 & 0.03 & 0.07 & 0.07 &~\,24.6 &~\,4.4 &~\,6.69 & a  \\ 
LP 707-58       & 01 18 15.97 & $-$12 53 59.6 & 2000.575 &~\,9.60 & 1.19 & 0.96 & 0.03 & 0.06 & 0.07 &~\,45.2 &~\,3.6 &~\,7.88 & a  \\ 
L 367-82        & 01 41 03.64 & $-$43 38 09.9 & 1998.923 &~\,9.90 & 1.03 & 0.87 & 0.04 & 0.06 & 0.08 &~\,23.7 &~\,2.6 &~\,6.77 & a  \\ 
LHS 6033        & 01 46 36.78 & $-$08 38 57.4 & 1998.710 & 10.28  & 1.42 & 0.85 & 0.03 & 0.09 & 0.12 &~\,70.1 & 14.2  &~\,9.51 & b  \\
G 271-177       & 01 53 45.45 & $-$06 03 02.1 & 1996.688 & 10.02  & 0.95 & 0.84 & 0.03 & 0.08 & 0.07 &~\,19.0 &~\,3.6 &~\,6.41 & a  \\ 
L 297-54        & 02 36 38.98 & $-$46 54 18.8 & 1998.929 &~\,9.53 & 0.93 & 0.91 & 0.02 & 0.07 & 0.07 &~\,25.8 &~\,3.4 &~\,6.59 & a  \\ 
LHS 1426        & 02 37 29.71 & $+$00 21 27.8 & 2000.838 & 12.10  & 1.51 & 0.90 & 0.03 & 0.07 & 0.08 &~\,40.2 &~\,4.3 & 10.12  & b  \\ 
LHS 1438        & 02 43 53.24 & $-$08 49 46.0 & 2000.899 &~\,9.83 & 1.07 & 0.84 & 0.03 & 0.08 & 0.07 &~\,42.2 &~\,3.4 &~\,7.96 & a  \\ 
LHS 17          & 02 46 14.97 & $-$04 59 21.5 & 2000.655 & 12.66  & 1.76 & 0.83 & 0.03 & 0.07 & 0.08 &~\,60.3 &~\,8.2 & 11.56  & b  \\ 
LP 771-21       & 02 48 40.98 & $-$16 51 21.9 & 2000.803 & 15.29  & 2.73 & 1.20 & 0.04 & 0.08 & 0.10 &~\,59.5 &~\,5.4 & 14.16  & c  \\ 
T* 831-161058   & 02 51 13.25 & $+$00 47 36.8 & 2000.901 & 16.51  & 2.74 & 1.19 & 0.08 & 0.11 & 0.14 &~\,20.5 &~\,2.2 & 13.07  & b  \\ 
T* 831-165166   & 02 51 42.68 & $-$01 02 05.6 & 2000.896 & 16.52  & 2.29 & 1.09 & 0.08 & 0.12 & 0.18 &~\,19.5 &~\,3.9 & 12.97  & c  \\ 
LP 994-59       & 03 09 27.87 & $-$42 28 50.7 & 1999.885 & 10.24  & 1.17 & 0.81 & 0.03 & 0.07 & 0.08 &~\,34.0 &~\,2.4 &~\,7.90 & a  \\ 
G 077-055       & 03 29 04.06 & $+$01 40 07.8 & 2000.759 &~\,9.99 & 1.05 & 0.80 & 0.03 & 0.08 & 0.09 &~\,21.2 &~\,3.6 &~\,6.62 & a  \\
Gl 145          & 03 32 55.83 & $-$44 42 07.1 & 1999.896 &~\,9.09 & 1.29 & 0.90 & 0.03 & 0.06 & 0.08 &~\,92.0 &~\,1.8 &~\,8.91 & a  \\ 
LHS 1565        & 03 35 59.61 & $-$44 30 45.5 & 1998.753 &~\,9.53 & 2.00 & 0.77 & 0.03 & 0.10 & 0.14 & 273.4  &~\,5.2 & 11.71  & e  \\
LP 944-20       & 03 39 35.26 & $-$35 25 43.6 & 2001.049 & 13.96  & 3.27 & 1.19 & 0.05 & 0.09 & 0.10 & 200.0  &~\,4.2 & 15.47  & c  \\ 
LHS 1604        & 03 51 00.03 & $-$00 52 44.6 & 1999.907 & 13.75  & 2.56 & 0.99 & 0.03 & 0.07 & 0.09 &~\,68.1 &~\,1.8 & 12.92  & b  \\ 
LHS 1832        & 06 10 59.85 & $-$65 12 20.3 & 1998.877 &~\,9.40 & 1.12 & 0.83 & 0.02 & 0.05 & 0.06 &~\,33.6 &~\,4.4 &~\,7.03 & a  \\
L 309-4         & 06 29 01.40 & $-$45 21 59.2 & 2000.145 &~\,9.12 & 0.98 & 0.80 & 0.03 & 0.08 & 0.07 &~\,30.2 &~\,1.5 &~\,6.52 & a  \\
LP 839-11       & 06 32 08.83 & $-$27 01 58.7 & 2001.085 &~\,9.70 & 1.06 & 0.88 & 0.03 & 0.07 & 0.07 &~\,37.0 &~\,2.6 &~\,7.54 & a  \\ 
LHS 1855        & 06 33 50.14 & $-$58 31 45.6 & 1996.066 &~\,9.46 & 1.58 & 0.82 & 0.02 & 0.07 & 0.08 &~\,61.3 &~\,1.8 &~\,8.40 & a  \\
G 108-024       & 06 44 13.94 & $-$00 55 31.5 & 1999.019 &~\,9.76 & 1.00 & 0.82 & 0.04 & 0.05 & 0.06 &~\,23.8 &~\,3.1 &~\,6.64 & a  \\
LHS 234         & 07 40 19.31 & $-$17 24 45.5 & 1999.192 & 12.36  & 2.20 & 0.85 & 0.02 & 0.05 & 0.07 & 112.4  &~\,2.7 & 12.61  & b  \\
HIP 39436       & 08 03 40.87 & $-$24 28 35.1 & 1998.964 &~\,9.63 & 0.94 & 0.74 & 0.03 & 0.05 & 0.11 &~\,22.2 &~\,3.1 &~\,6.36 & a  \\ 
L 98-45         & 08 19 16.07 & $-$67 48 14.3 & 1996.964 & 10.16  & 1.01 & 0.86 & 0.02 & 0.05 & 0.06 &~\,26.9 &~\,2.4 &~\,7.31 & a  \\ 
LP 665-21       & 08 31 21.75 & $-$06 02 01.4 & 1996.038 &~\,9.26 & 1.21 & 0.89 & 0.03 & 0.05 & 0.06 &~\,46.0 &~\,6.4 &~\,7.57 & a  \\ 
LHS 6149        & 08 34 25.91 & $-$01 08 39.3 & 2000.022 & 10.11  & 1.34 & 0.69 & 0.02 & 0.06 & 0.06 &~\,73.4 &~\,9.6 &~\,9.44 & b  \\ 
LP 98-62        & 08 41 32.69 & $-$68 25 40.6 & 1999.238 &~\,9.24 & 0.99 & 0.88 & 0.02 & 0.04 & 0.06 &~\,32.1 &~\,1.6 &~\,6.77 & a  \\
LHS 2145        & 09 28 53.34 & $-$07 22 16.1 & 2000.326 &~\,9.69 & 1.28 & 0.76 & 0.03 & 0.07 & 0.07 &~\,58.2 &~\,4.1 &~\,8.51 & a  \\ 
LHS 2264        & 10 26 07.80 & $-$17 58 43.5 & 1996.044 &~\,9.60 & 1.02 & 0.75 & 0.02 & 0.06 & 0.07 &~\,29.1 &~\,2.3 &~\,6.92 & a  \\ 
LHS 292         & 10 48 12.64 & $-$11 20 09.8 & 2000.200 & 11.25  & 2.30 & 0.98 & 0.03 & 0.07 & 0.06 & 220.3  &~\,3.6 & 12.97  & b  \\ 
DENIS 1048-39   & 10 48 14.42 & $-$39 56 08.2 & 2001.359 & 12.64  & 3.00 & 1.13 & 0.03 & 0.07 & 0.07 & 192.0  & 37.0  & 14.06  & d  \\ 
LP 672-4        & 11 09 12.28 & $-$04 36 24.9 & 1999.378 &~\,9.49 & 1.40 & 0.77 & 0.03 & 0.06 & 0.06 &~\,39.9 &~\,2.4 &~\,7.49 & a  \\ 
LHS 2397a       & 11 21 49.21 & $-$13 13 08.3 & 2000.501 & 14.97  & 3.09 & 1.25 & 0.10 & 0.11 & 0.08 &~\,70.0 &~\,2.1 & 14.09  & b  \\ 
LP 793-34$^{+}$ & 11 45 35.40 & $-$20 21 05.2 & 2000.241 & 13.84  & 2.15 & 0.88 & 0.05 & 0.05 & 0.10 &~\,49.6 &~\,3.6 & 12.32  & a  \\ 
LHS 314         & 11 46 42.93 & $-$14 00 51.8 & 2000.205 &~\,9.33 & 1.30 & 0.96 & 0.03 & 0.07 & 0.07 &~\,49.0 &~\,2.9 &~\,7.78 & a  \\ 
LHS 2475        & 11 55 07.44 & $+$00 58 25.9 & 1996.208 &~\,9.39 & 1.18 & 0.87 & 0.03 & 0.05 & 0.07 &~\,35.8 &~\,3.2 &~\,7.16 & a  \\ 
    \noalign{\smallskip}
    \hline 
   \end{tabular}
  $$
\end{table*}
\begin{table*}
\setcounter{table}{0}
   \caption{Continued}
   \label{table_cali2}
  $$
   \begin{tabular}{llllllllllllll}
   \hline
   \noalign{\smallskip}
Stars &$\alpha_{\rm 2000}$&$\delta_{\rm 2000}$&DENIS& I  &I$-$J &J$-$K&errI&errJ&errK&$\pi$&err$\pi$&$M_{\rm I}$& Ref \\ 
      &                   &                   &Epoch&    &      &     &    &    &    & mas & mas&          &     \\ 
   \hline
   \noalign{\smallskip}  
LHS 2477        & 11 55 49.22 & $-$38 16 49.7 & 2001.104 &~\,9.88 & 1.12 & 0.75 & 0.03 & 0.07 & 0.09 &~\,42.8 &~\,3.0 &~\,8.04 & a  \\ 
LHS 2509        & 12 04 36.61 & $-$38 16 25.2 & 2000.233 &~\,9.74 & 1.15 & 0.87 & 0.04 & 0.08 & 0.05 &~\,37.3 &~\,4.9 &~\,7.60 & a  \\ 
LP 794-30       & 12 11 11.78 & $-$19 57 38.1 & 1999.148 &~\,9.48 & 1.57 & 0.89 & 0.03 & 0.05 & 0.08 &~\,78.1 &~\,3.1 &~\,8.94 & a  \\ 
LP 852-57       & 12 13 32.93 & $-$25 55 24.5 & 1999.153 &~\,9.48 & 1.08 & 0.84 & 0.03 & 0.06 & 0.08 &~\,42.1 &~\,2.5 &~\,7.60 & a  \\ 
LHS 2587        & 12 36 49.29 & $-$76 57 17.8 & 1998.197 &~\,9.37 & 0.96 & 0.85 & 0.02 & 0.07 & 0.07 &~\,27.8 &~\,1.7 &~\,6.59 & a  \\ 
LHS 2595        & 12 38 47.34 & $-$04 19 17.0 & 1999.414 & 10.80  & 1.46 & 0.86 & 0.02 & 0.06 & 0.08 &~\,50.7 &~\,3.1 &~\,9.33 & b  \\
LP 617-37       & 13 20 24.96 & $-$01 39 26.3 & 1999.211 &~\,9.57 & 1.23 & 0.85 & 0.03 & 0.06 & 0.09 &~\,48.2 &~\,2.9 &~\,7.98 & a  \\ 
LP 855-14       & 13 27 53.95 & $-$26 57 01.8 & 2001.151 &~\,9.60 & 1.25 & 0.88 & 0.03 & 0.08 & 0.09 &~\,48.0 &~\,2.9 &~\,8.01 & a  \\ 
LHS 2770        & 13 38 24.73 & $-$02 51 51.9 & 1999.279 & 12.54  & 1.43 & 0.95 & 0.04 & 0.07 & 0.07 &~\,26.3 &~\,6.7 &~\,9.64 & b  \\ 
LP 912-26       & 13 53 19.76 & $-$30 46 37.6 & 2000.364 & 10.02  & 1.06 & 0.97 & 0.03 & 0.06 & 0.09 &~\,27.0 &~\,3.8 &~\,7.18 & a  \\ 
LHS 2876        & 14 12 12.17 & $-$00 35 16.2 & 1999.444 & 15.59  & 2.50 & 1.01 & 0.05 & 0.08 & 0.08 &~\,32.7 &~\,4.1 & 13.16  & c  \\ 
T* 868-110639   & 15 10 16.86 & $-$02 41 07.4 & 1999.384 & 15.73  & 3.09 & 1.32 & 0.05 & 0.07 & 0.09 &~\,57.5 &~\,1.9 & 14.53  & b  \\ 
LHS 392         & 15 11 50.60 & $-$10 14 17.8 & 2000.277 & 11.24  & 1.47 & 0.90 & 0.04 & 0.09 & 0.12 &~\,67.4 &~\,3.1 & 10.38  & b  \\ 
LP 915-16       & 15 17 21.16 & $-$27 59 49.8 & 1996.422 &~\,9.58 & 1.27 & 0.91 & 0.02 & 0.08 & 0.12 &~\,41.2 &~\,3.7 &~\,7.66 & a  \\ 
LHS 3092        & 15 36 34.53 & $-$37 54 22.3 & 1999.211 &~\,9.91 & 1.47 & 0.79 & 0.03 & 0.05 & 0.06 &~\,81.6 & 13.7  &~\,9.47 & b  \\ 
LHS 3093        & 15 36 58.69 & $-$14 08 00.7 & 1998.373 & 10.02  & 1.63 & 0.88 & 0.02 & 0.05 & 0.05 &~\,74.9 &~\,3.8 &~\,9.39 & b  \\ 
LP 336-71       & 15 49 38.34 & $-$47 36 33.8 & 1999.290 &~\,9.46 & 1.13 & 0.88 & 0.03 & 0.05 & 0.06 &~\,37.5 &~\,2.6 &~\,7.33 & a  \\ 
LP 744-46       & 16 02 35.07 & $-$14 38 36.5 & 1996.359 &~\,9.71 & 1.16 & 0.85 & 0.03 & 0.06 & 0.09 &~\,31.4 &~\,4.1 &~\,7.19 & a  \\ 
LHS 412         & 16 08 15.03 & $-$10 26 11.7 & 1998.323 & 11.78  & 1.51 & 0.81 & 0.03 & 0.07 & 0.09 &~\,47.1 &~\,2.7 & 10.15  & b  \\
LHS 3185        & 16 22 40.97 & $-$48 39 19.7 & 1999.570 &~\,9.70 & 1.26 & 0.78 & 0.04 & 0.08 & 0.05 &~\,41.0 &~\,3.7 &~\,7.76 & a  \\ 
LP 625-34       & 16 40 05.98 & $+$00 42 19.3 & 1999.625 & 10.67  & 1.57 & 0.84 & 0.02 & 0.06 & 0.06 &~\,89.0 &~\,2.3 & 10.42  & b  \\ 
LHS 3242        & 16 48 24.40 & $-$72 58 33.9 & 2000.537 &~\,9.27 & 1.22 & 0.80 & 0.03 & 0.08 & 0.06 &~\,62.7 &~\,1.9 &~\,8.25 & a  \\ 
LHS 3272        & 17 13 40.46 & $-$08 25 14.6 & 2000.573 &~\,9.54 & 1.39 & 0.83 & 0.04 & 0.07 & 0.27 &~\,52.8 &~\,4.2 &~\,8.15 & a  \\ 
HIP 86938       & 17 45 53.36 & $-$13 18 22.1 & 2000.551 & 10.14  & 1.06 & 0.81 & 0.04 & 0.08 & 0.06 &~\,26.9 &~\,3.8 &~\,7.29 & a  \\ 
HIP 91644       & 18 41 19.73 & $-$60 25 47.4 & 2000.381 &~\,9.45 & 1.01 & 0.76 & 0.02 & 0.06 & 0.09 &~\,27.5 &~\,2.5 &~\,6.65 & a  \\ 
LHS 3421        & 18 52 52.30 & $-$57 07 38.1 & 2000.773 &~\,9.84 & 1.35 & 0.89 & 0.03 & 0.07 & 0.07 &~\,37.5 &~\,3.8 &~\,7.71 & a  \\ 
L 850-62        & 19 03 16.64 & $-$13 34 05.4 & 2000.573 & 11.93  & 1.57 & 0.81 & 0.03 & 0.07 & 0.07 &~\,52.4 &~\,3.8 & 10.53  & b  \\ 
LTT 7598        & 19 12 25.27 & $-$55 52 07.6 & 1999.512 &~\,9.47 & 1.19 & 0.87 & 0.02 & 0.05 & 0.06 &~\,50.0 &~\,2.5 &~\,7.97 & a  \\ 
LP 635-46       & 20 43 41.32 & $-$00 10 41.3 & 1999.605 &~\,9.54 & 1.02 & 0.88 & 0.02 & 0.06 & 0.07 &~\,38.4 &~\,3.1 &~\,7.46 & a  \\ 
LP 211-96       & 20 59 51.36 & $-$58 45 31.1 & 2001.359 &~\,9.71 & 1.14 & 0.85 & 0.04 & 0.08 & 0.07 &~\,32.0 &~\,2.9 &~\,7.24 & a  \\ 
LHS 3639        & 21 11 49.56 & $-$43 36 48.8 & 1999.540 &~\,9.59 & 1.14 & 0.77 & 0.08 & 0.08 & 0.07 &~\,69.8 &~\,4.2 &~\,8.81 & a  \\ 
LHS 3666        & 21 24 18.32 & $-$46 41 35.3 & 1999.559 & 10.20  & 1.20 & 0.86 & 0.03 & 0.06 & 0.07 &~\,37.2 &~\,4.8 &~\,8.05 & a  \\ 
HB 2124-4228    & 21 27 26.12 & $-$42 15 18.1 & 1998.652 & 16.02  & 2.47 & 1.45 & 0.06 & 0.13 & 0.16 &~\,28.0 &~\,6.2 & 13.26  & c  \\
HIP 106043      & 21 28 44.42 & $-$47 15 42.2 & 1998.501 & 10.36  & 1.04 & 0.98 & 0.05 & 0.11 & 0.12 &~\,26.7 &~\,4.0 &~\,7.49 & a  \\
LHS 513         & 21 39 00.66 & $-$24 09 26.7 & 1996.638 & 10.68  & 1.55 & 0.74 & 0.04 & 0.06 & 0.08 &~\,73.3 & 12.0  & 10.01  & b  \\ 
LHS 5374        & 21 54 45.25 & $-$46 59 34.5 & 2000.605 &~\,9.73 & 1.32 & 0.88 & 0.03 & 0.08 & 0.08 &~\,66.1 &~\,3.3 &~\,8.83 & a  \\
HIP 108523      & 21 59 08.30 & $-$46 45 47.3 & 1998.679 &~\,9.74 & 1.10 & 1.03 & 0.03 & 0.11 & 0.11 &~\,37.8 &~\,3.8 &~\,7.63 & a  \\
LP 283-3        & 22 03 27.19 & $-$50 38 39.2 & 2000.512 &~\,9.83 & 1.14 & 0.87 & 0.04 & 0.07 & 0.14 &~\,45.9 &~\,8.3 &~\,8.14 & a  \\
LHS 3776        & 22 13 42.90 & $-$17 41 08.8 & 2000.504 & 10.65  & 1.70 & 0.84 & 0.08 & 0.06 & 0.06 &~\,96.0 &~\,3.9 & 10.56  & b  \\ 
T* 890-60235    & 22 23 05.56 & $+$00 30 11.1 & 1999.614 & 16.62  & 2.43 & 1.18 & 0.07 & 0.10 & 0.14 &~\,19.4 &~\,2.2 & 13.06  & c  \\ 
HIP 110655      & 22 25 02.83 & $-$33 12 16.2 & 2000.458 &~\,9.02 & 0.92 & 0.75 & 0.04 & 0.08 & 0.08 &~\,30.7 &~\,5.2 &~\,6.46 & a  \\
LHS 523         & 22 28 54.38 & $-$13 25 17.8 & 1998.729 & 12.87  & 2.19 & 0.89 & 0.04 & 0.11 & 0.13 &~\,88.8 &~\,4.9 & 12.61  & b  \\
LHS 526         & 22 34 53.61 & $-$01 04 58.0 & 1998.723 & 11.89  & 1.47 & 1.04 & 0.03 & 0.09 & 0.17 &~\,42.5 &~\,3.7 & 10.03  & b  \\
LHS 3850        & 22 46 26.28 & $-$06 39 25.0 & 1998.474 & 12.62  & 1.94 & 0.80 & 0.02 & 0.10 & 0.12 &~\,53.3 &~\,4.6 & 11.25  & b  \\ 
HIP 114252      & 23 08 19.55 & $-$15 24 35.8 & 1999.466 &~\,9.17 & 1.15 & 0.93 & 0.02 & 0.07 & 0.06 &~\,45.8 &~\,2.7 &~\,7.47 & a  \\
G 157-52        & 23 21 11.25 & $-$01 35 44.9 & 2000.578 &~\,9.77 & 1.11 & 0.82 & 0.03 & 0.07 & 0.08 &~\,37.0 &~\,3.7 &~\,7.61 & a  \\ 
LHS 546         & 23 35 10.45 & $-$02 23 19.9 & 1999.696 & 11.01  & 1.87 & 0.97 & 0.03 & 0.07 & 0.07 & 138.3  &~\,3.5 & 11.71  & b  \\ 
HIP 118180      & 23 58 22.03 & $-$53 48 33.6 & 1999.874 &~\,9.21 & 0.97 & 0.68 & 0.03 & 0.07 & 0.07 &~\,29.7 &~\,3.0 &~\,6.57 & a  \\
   \noalign{\smallskip}
   \hline 
   \end{tabular}
  $$
  \begin{list}{}{}
  \item[T*] : TVLM 
  \item[$^{+}$]: Hipparcos, for LP 793-33 
  \item[*] : BRI 0021-02. This object is also listed in the NLTT, as 
LP~585-86. That name is clearly an NLTT typo: another star (the much brighter 
HIP~3061) bears the same name, with coordinates that are consistent
with the LP numbering sequence. The NLTT proper motion is on the other hand
valid: 0.212, 320~degrees.

Columns 1, 2, 3 \& 4: Object name, DENIS Position for equinox J2000 at 
DENIS epoch, and DENIS epoch.

Columns 5, 6 \& 7; and 8, 9 \& 10: DENIS I-magnitude and colours; and 
associated standard errors.  

Columns 11 \& 12: Trigonometric parallax and its standard error.

Column 13:  M$_{\rm I}$ absolute magnitude, calculated from DENIS I-magnitude
and parallax.

Column 14: Parallax reference: (a) HIP; (b) GCTP; (c) Tinney et al. 
(\cite{tinney95}) and Tinney (\cite{tinney96}); (d) Deacon \& Hambly (\cite{deacon}); (e) Henry et al. (\cite{henry97}).
  \end{list}
\end{table*}
The DEep Near Infrared Survey (DENIS) (Epchtein \cite{epchtein97})
systematically surveyed the southern sky in two near-infrared 
($J$ and $K_{\rm S}$) and one optical ($I$) band. Its extensive sky
coverage, broad wavelength baseline, and moderately deep exposures
($I$=18.5, $J$=16, $K_{\rm S}$=13.5) make it a very efficient tool 
at identifying faint and cool nearby stars.

In Paper I, we estimated distances to potential DENIS red dwarfs using the 
Cousins-CIT ($I_{\rm C}-J_{\rm CIT}$, $M_{\rm I}$) relation for M dwarfs
of Delfosse (\cite{delfosseb}).
We also noted that for red stars the DENIS photometric system and the standard
Cousins-CIT system differ by $\sim$~0.1~mag for the $K$ band, but by less 
than 0.05~mag for the $I$ and $J$ bands (Delfosse \cite{delfosseb}). The
Delfosse (\cite{delfosseb}) $I-J$ relation therefore applies reasonably well
to DENIS photometry, but with progress in the DENIS data reduction it 
has now become possible, and preferable, to directly calibrate a DENIS 
colour-magnitude relation.
Of the three colours that can be formed from DENIS photometry, $J-K$ is a
very poor spectral type diagnostic for M dwarfs, while $I-J$ and $I-K$ are
both excellent. From a practical point of view, DENIS is significantly
more sensitive to M dwarfs at $J$ than at $K$. We therefore chose to 
calibrate the ($I-J$, $M_{\rm I}$) relation.

We searched the following  trigonometric parallax catalogues
for reference M dwarfs 
with a DENIS counterpart fainter than the $I$ saturation limit
of $I=9$ and with $I-J>1.0$:
\begin{itemize}
 \item[\bf\Huge{.}]  the Hipparcos catalogue (ESA \cite{esa}) for 63 relatively
        bright stars. As the limiting magnitude of the HIP catalogue is 
        V$\sim 12.0$, it contains few very red dwarfs. 
 \item[\bf\Huge{.}]  the GCTP catalogue (van Altena et al. \cite{vanaltena}) 
     for 29 mostly fainter stars.
 \item[\bf\Huge{.}]  6 faint stars from Tinney et al. (\cite{tinney95}), 
          Tinney (\cite{tinney96});
      one from Henry et al. (\cite{henry97}) 
      and one late M dwarf from Deacon \& Hambly (\cite{deacon}).
\end{itemize}
We excluded known doubles as well as large amplitude variables, but had
to accept  a number of low amplitude flare stars, with peak visible 
light amplitude of 0.1 to 0.3~mag.

We did not correct the resulting absolute magnitudes for the Lutz-Kelker 
bias, since the complex selection pedigree of our sample makes a quantitative
analysis of that bias almost impossible. Arenou \& Luri (\cite{arenou}) 
conclude that it is preferable to apply no correction in such cases.
The errors on the parallaxes are fortunately small, so that neglecting
that correction does not appreciably contribute to the overall errors.

Fig.~\ref{fig_col_mag} shows the resulting ($I-J$, $M_{\rm I}$) plot, 
and the corresponding $4^{th}$ order polynomial fit:
\begin{eqnarray}
 M_{\rm I} & = & a_{0}+a_{1}(I-J)+a_{2}(I-J)^{2}+a_{3}(I-J)^{3} \nonumber \\ 
           &   & +~a_{4}(I-J)^{4} \label{eq1}
\end{eqnarray}
where $a_{0}=11.370$, $a_{1}=-19.175$, $a_{2}=21.587$, $a_{3}=-7.877$, 
$a_{4}=0.9710$, valid for $0.9~\leq~I-J~\leq~3.1$. 

Reid \& Cruz (\cite{reid}) established a similar relation for the
Cousins/CIT colours, which only differ slightly from the DENIS colours.
That relation is illustrated on Fig.~\ref{fig_col_mag}, together with
the theoretical prediction of Baraffe (\cite{baraffe}). 
In the $[1.7,3.1]$ range we could collect 22 data points, significantly 
more than the 14 objects that we count on Fig.~11 of Reid \& Cruz 
(\cite{reid}).

In the $[1.0,1.4]$ interval the three curves are very close, but
they then  disagree over the intermediate $[1.4,1.8]$ region where the 
colour-luminosity relation steepens considerably. 
Reid \& Cruz (\cite{reid}) choose to describe this difficult region by
discontinuities at $I-J=1.45$ and $I-J=1.65$, with a constant value
with large error bars  ($M_{\rm I}$=10.2$\pm$0.7) used in-between. The
discontinuities clearly are non-physical, but our polynomial fit, just
as clearly, runs the risk of smoothing out a steeper intrinsic slope
which could reflect a real physical change or transition in the 
stellar structure.

Which of the two description is preferable largely rests on a small
group of stars in this colour range: LHS~1855 (Gl~238), LP~794-30, 
LHS~3093 (Gl~592), G~70-22, and to a lesser extend, LHS~3850 (GJ~4294). 
If those stars are single, the polynomial fit is clearly preferable 
to the Reid \& Cruz prescription, but an alternative hypothesis is that 
they are photometric binaries. Of the five, two have been examined
for companions (LHS 1855, Scholz et al. \cite{scholz00}; LHS 3093, Skrutskie 
et al. \cite{skrutskie}) and found single, but only with seeing-limited
resolution. LP 794-30 has a known companion, but at 85$''$, outside any
photometric diaphragm. We observed both G~70-22 and LHS~3850 with adaptive
optics at CFHT, and found the former resolved with $\Delta$(K)=1.5 at a 
separation of 0.8''. For now we lack objective reasons to excise the other 
4 stars from the list and have thus left them in, 
but we did add them to the observing lists of our adaptive optics 
and radial velocity programs (Delfosse et al. \cite{delfosse99b}).

In the $I-J$ range of primary interest here ([2.0,3.0]) all three 
relations again agree well with the data, as seen in Fig.~\ref{residu} 
which plots the residuals  of the observed data points from the three fits. 
Over the $[1.9,3.1]$ range the rms dispersion of the data around our 
fit is 0.26, corresponding to a 12\% error on distances; it is respectively 
0.30 and 0.31 for the Baraffe et al. (\cite{baraffe}) and Reid \& Cruz 
(\cite{reid}) relations. Over the $[2.0,3.0]$ range the Eq.~(\ref{eq1}) 
polynomial is therefore a small but significant improvement, and we use
it for the reminder of this paper. 


\section{ Sample selection}
\subsection{Star selection from the DENIS survey}
\begin{figure}
\psfig{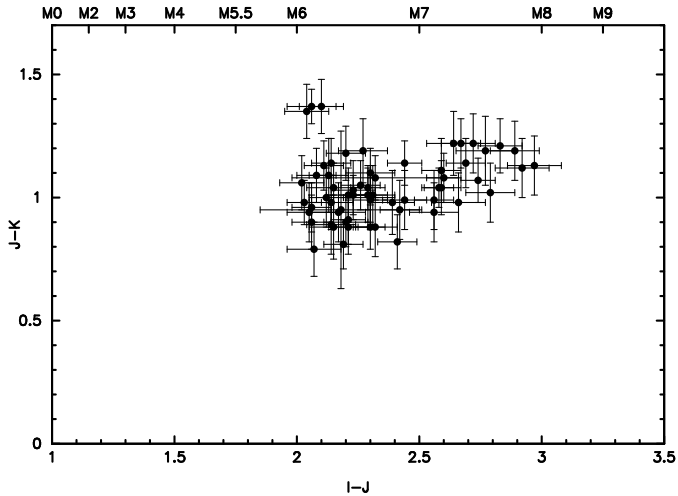}
\caption{ DENIS colour-colour diagram for all 62 late-M dwarf candidates
detected in the 5700 square degrees (Paper I and this paper) so far 
examined in DENIS. Stars selected from the NLTT outside this area
are not shown. The (indicative) spectral type
labels on the top axis are adopted from Leggett (\cite{leggett})}.
\label{ij_jk}
\end{figure}
We systematically search the DENIS database (available at the Paris Data 
Analysis Center, PDAC) for potential members of the solar neighbourhood,
with simple and well defined criteria. Specifically, we start by selecting 
all high galactic latitude DENIS sources ($|b_{II}|~{\geq}~30\degr$) that
are redder than $I-J=1.0$ (approximately the colour of an M0 dwarf, 
Leggett \cite{leggett}). 
We then compute photometric distances to retain stars with 
$D_{\rm phot} < 30$~pc. We used the Paper~I colour-magnitude relation
for this selection since the colour-relation presented above was 
not yet available when we queried the DENIS database, but later
recomputed all distances with the new relation.

When the search program was last run in mid-2001, 
5700~square degrees (slightly over half of the southern 
high galactic latitude sky) were available in the database 
(Delfosse \& Forveille 
\cite{delfossed}). 2100 of those 5700 square degrees had been considered 
in Paper~I and are reanalysed here with slightly improved tools, and 3600
square degrees are new. The number of potential early-M dwarfs (M0 to 
M6, $1.0 \leq I-J \leq 2.0$) with photometric distances within 30~pc is 
significantly larger ($\sim$5000) than the total population expected for
the sampled volume ($\sim$1400, Henry et al. \cite{henry02}), and therefore
must be dominated by contamination from distant M-giants with similar 
colours. Its analysis will require considerable follow-up, 
which is beyond the scope 
of the present work. Very late-M and L dwarfs ($I-J \geq 3.0$) will be 
considered in a forthcoming paper (Delfosse et al., in preparation).
As in Paper~I, we thus restrict the present analysis to stars in the 
$2.0 \leq I-J \leq 3.0$ colour interval, or approximately 
to spectral types M6 to M8. 

At that stage in the selection, the candidate sample contains approximately 
equal numbers of nearby dwarfs and distant giants. A cut in the 
$I-J$/$J-K$ colour-colour diagram  (Fig.~\ref{ij_jk}) rejects a sizeable
fraction of the giants, but the DENIS photometry is not sufficiently accurate
to eliminate all of them without losing some dwarfs. That step needs proper
motion information, and is discussed in Sec.~5. 

\subsection{Searching in the NLTT catalogue}
To extend our search to lower galactic latitudes, we turned to known high
proper motion stars, and looked for faint NLTT (Luyten \cite{luytenb}) stars
with DENIS colours and magnitudes compatible with a nearby late M-dwarf.
As the brighter NLTT stars have usually been better characterised, we
restricted that search to NLTT stars fainter than $R_{\rm Luyten}=14.0$ and 
redder than $(B-R)_{\rm Luyten}=1.5$ (approximately later than M1, Leggett 
\cite{leggett}). The resulting 7424 stars were searched
for in the 14,000 square degrees of DENIS data that are presently 
processed and available on-line at PDAC. This cross-identification is 
made somewhat difficult by the interplay of crowded fields at low galactic
latitudes with the often poor coordinates of the southern stars in the NLTT 
catalogue. We therefore expect to have missed some significant fraction
of the true matches. These NLTT stars were then handled as those extracted
directly from DENIS, except that they obviously were ignored during the
statistical analysis of the DENIS sample. We present here 15 candidates
matching  our previous criteria ($2.0 \leq I-J \leq 3.0$; and photometric 
distance within 30~pc).

%
%
%
\begin{table*}
   \caption[A]{{\bf{a)}} Proper motions of the 24 high-PM 
($\mu~>~0.1\arcsec$yr$^{\rm -1}$) late-M dwarfs selected in the 3600 square degrees}
    \label{pma}
  $$
   \begin{tabular}{lllll}
   \hline 
   \hline
   \noalign{\smallskip}
DENIS Name&$\mu_{\rm \alpha}$&$\mu_{\rm \delta}$&$\mu_{\rm total}$&$\mu_{\rm _L}$ \\
          &[$\arcsec$yr$^{-1}$]&[$\arcsec$yr$^{-1}$]&[$\arcsec$yr$^{-1}$]&[$\arcsec$yr$^{-1}$] \\
    \hline
    \noalign{\smallskip}
 J0020231$-$234605      & +0.322 & $-$0.066 & 0.329 & 0.370 \\
 J0103119$-$535143*     &$-$0.094& $-$0.218 & 0.238 & ...   \\
 J0120491$-$074103*     &$-$0.013& $-$0.115 & 0.116 & ...   \\
 J0144318$-$460432*     & +0.117 & $-$0.049 & 0.127 & ...   \\
 J0218579$-$061749      & +0.367 & $-$0.097 & 0.379 & 0.375 \\
 J0235495$-$071121*     & +0.284 & +0.093   & 0.299 & ...   \\
 J0306115$-$364753*     &$-$0.181& $-$0.700 & 0.723 & ...   \\
 J0320588$-$552015*     & +0.302 & +0.259   & 0.398 & ...   \\
 J0351000$-$005244      & +0.035 & $-$0.475 & 0.477 & 0.525 \\
 J0517377$-$334903*     & +0.464 & $-$0.342 & 0.576 & ...   \\
 J1006319$-$165326      &$-$0.318& +0.181   & 0.366 & 0.391 \\
 J1021513$-$032309      & +0.202 & $-$0.147 & 0.249 & 0.269 \\
 J1048126$-$112009      & +0.604 & $-$1.521 & 1.637 & 1.644 \\
 J1106569$-$124402      &$-$0.314& +0.001   & 0.314 & 0.355 \\
 J1141440$-$223215*     &$-$0.141& +0.400   & 0.424 & ...   \\
 J1145354$-$202105      & +0.149 & +0.063   & 0.161 & 0.186 \\
 J1147421$+$001506      &$-$0.262&$-$0.083  & 0.275 & 0.303 \\
 J1155429$-$222458      &$-$0.377&$-$0.185  & 0.420 & 0.412 \\
 J1201421$-$273746      &$-$0.289&$-$0.187  & 0.344 & 0.302 \\
 J1250526$-$212113*     & +0.441 &$-$0.340  & 0.557 & ...   \\ 
 J1610584$-$063132      &$-$0.051&$-$0.180  & 0.187 & 0.229 \\
 J2132297$-$051158      & +0.109 &$-$0.337  & 0.354 & 0.350 \\
 J2205357$-$110428      &$-$0.271&$-$0.166  & 0.318 & 0.339 \\
 J2337383$-$125027      & +0.205 &$-$0.312  & 0.373 & 0.365 \\
    \noalign{\smallskip}
    \hline 
   \end{tabular}
  $$
  \begin{list}{}{}
  \item[$^{*}$] Not previously known as a high-PM star.
  
  Column 1: Object name.

  Columns 2, 3 \& 4: $\mu_{\rm \alpha}$, $\mu_{\rm \delta}$, $\mu_{\rm total}$, 
    our measurements, in arc-sec.yr$^{\rm -1}$. 
  
  Column 5: Total proper motion from Luyten (\cite{luytena}, 
   \cite{luytenb}), when available.
  \end{list}
\end{table*}
\begin{table*}
\setcounter{table}{1}
   \caption{{\bf{b)}} Proper motions of the 15 high-PM 
($\mu~>~0.1\arcsec$yr$^{\rm -1}$) late-M dwarfs initially selected 
from the NLTT (same columns as Table~\ref{pma}a)}
    \label{pmb}
  $$
   \begin{tabular}{lllll}
   \hline 
   \hline
   \noalign{\smallskip}
DENIS Name&$\mu_{\rm \alpha}$&$\mu_{\rm \delta}$&$\mu_{\rm total}$&$\mu_{\rm _L}$ \\
          & [$\arcsec$yr$^{-1}$]&[$\arcsec$yr$^{-1}$]&[$\arcsec$yr$^{-1}$]&[$\arcsec$yr$^{-1}$] \\
    \hline
    \noalign{\smallskip}
 J0002061$+$011536& +0.474 & +0.068   & 0.479 & 0.445 \\
 J0410480$-$125142&$-$0.117& $-$0.382 & 0.400 & 0.426 \\
 J0440231$-$053009& +0.313 & +0.101   & 0.329 & 0.243 \\
 J0520293$-$231848& +0.222 & +0.250   & 0.334 & 0.334 \\
 J0931223$-$171742&$-$0.286& $-$0.131 & 0.315 & 0.296 \\
 J1346460$-$314925&$-$0.336& +0.158   & 0.372 & 0.371 \\
 J1504161$-$235556&$-$0.317&$-$0.078  & 0.326 & 0.322 \\
 J1546115$-$251405&$-$0.218&$-$0.310  & 0.379 & 0.377 \\
 J1552446$-$262313& +0.227 &$-$0.475  & 0.526 & 0.492 \\
 J1553571$-$231152&$-$0.112&$-$0.281  & 0.303 & 0.299 \\
 J1625503$-$240008&$-$0.158&$-$0.026  & 0.161 & 0.184 \\
 J1641430$-$235948&$-$0.111&$-$0.185  & 0.216 & 0.212 \\
 J1645282$-$011228& +0.013 &$-$0.220  & 0.220 & 0.231 \\
 J1917045$-$301920& +0.191 &$-$0.207  & 0.281 & 0.212 \\
 J2151270$-$012713& +0.220 & +0.023   & 0.221 & 0.223 \\
    \noalign{\smallskip}
    \hline 
   \end{tabular}
  $$
  \begin{list}{}{}
  \item[] 
  \end{list}
\end{table*}
\begin{table*}
\setcounter{table}{1}
   \caption{{\bf{c)}} Proper motions of 11 low-PM ($\mu < 0.1\arcsec$yr$^{\rm -1}$) 
       probable late-M dwarfs found in the 5700 square degrees}
    \label{low_pm}
  $$
   \begin{tabular}{llllll}
   \hline 
   \hline
   \noalign{\smallskip}
DENIS Name         &$\mu_{\rm \alpha}$&$\mu_{\rm \delta}$& err~$\mu_{\rm \alpha}$& err~$\mu_{\rm \delta}$& $\mu_{\rm total}$ \\
      & [mas.yr$^{-1}$] & [mas.yr$^{-1}$] & [mas.yr$^{-1}$] & [mas.yr$^{-1}$] & [mas.yr$^{-1}$]    \\
    \hline
    \noalign{\smallskip}     
J0013093$-$002551  &  $+$97   & $+$~\,4 &  25   &  25   &  97      \\
J0100021$-$615627  &  $+$78   & $-$41   &  21   &  21   &  88      \\
J0436278$-$411446  &  $+$22   & $+$~\,4 &  18   &  18   &  22      \\
J0518113$-$310153  &  $+$41   & $-$~\,5 &  10   &  10   &  41      \\
J1236396$-$172216  &  $+$14   & $-$60   &  20   &  20   &  62      \\
J1538317$-$103850  &  $-$~\,8 & $-$18   & ~\,9  & ~\,9  &  20      \\
J1552237$-$033520  &  $-$~\,8 & $-$30   & ~\,9  & ~\,9  &  31      \\
J1553186$-$025919  &  $+$14   & $-$24   & ~\,8  & ~\,8  &  28      \\
J2022480$-$564556  &  $-$~\,1 & $-$84   &  17   &  17   &  84      \\
J2206227$-$204706  &  $+$28   & $-$57   &  29   &  29   &  64      \\
J2226443$-$750342  &  $+$48   & $+$14   &  19   &  19   &  50      \\
    \noalign{\smallskip}
    \hline 
   \end{tabular}
  $$
  \begin{list}{}{}
  \item[]  
  \end{list}
\end{table*}

\section{Proper motions and B, R magnitudes}
We searched for plates containing the dwarf candidates
in the collection of the Centre d'Analyse des Images (CAI, 
{\footnotesize http://www.cai-mama.obspm.fr/}):
POSS I ($-30\degr<\delta<0\degr$),
SRC-J ($-90\degr<\delta<0\degr$), SRC-R($-17\degr<\delta<0\degr$) and
ESO-R ( $\delta<-17\degr$), depending on the declination.
We then used the MAMA microdensitometer (Berger et al. \cite{berger}) at 
CAI to digitize the survey plates, and analysed the resulting images 
with SExtractor (Bertin \& Arnouts \cite{bertin}). 
We calibrated these measurements using the ACT (Urban et al. \cite{urban}) 
and GSPC-2 (Postman et al. 
\cite{postman}, Bucciarelli et al. \cite{bucciarelli}) catalogues,
as respectively astrometric and photometric references.

A least-square fit to the positions at the 3 to 4 available epochs
(including the DENIS survey epoch), determines absolute proper motion.
The time baseline spans 13 to 49 years, and results in proper motion
standard errors of 29 to 7~mas/year.
The photometric standard errors are  
$\pm 0.3$ mag for $B$ and $\pm 0.2$ mag for $R$. 
Tables~\ref{pma}a \&~\ref{low_pm}c
respectively list the proper motion determinations for 24 
high proper motions (high-PM, $\mu > 0.1\arcsec$yr$^{-1}$)
in the 3600 square degrees and 11 lower proper motions 
(low-PM, $\mu < 0.1\arcsec$yr$^{-1}$)
in the full 5700 square degrees.
Table~\ref{pmb}b lists the proper motions for 15 high-PM candidates 
initially selected from the NLTT catalog.

For some bright low-PM objects, we used $B$ and $R$ magnitudes available in 
the USNO-A2.0 catalogue (Monet at al. \cite{monet}), as well as
more accurate proper motions from the UCAC1 (Zacharias et al. 
\cite{zacharias}) \& Tycho-2 (H\o g et al. \cite{hog}) catalogues. 
\begin{figure}
\psfig{height=6.5cm,file=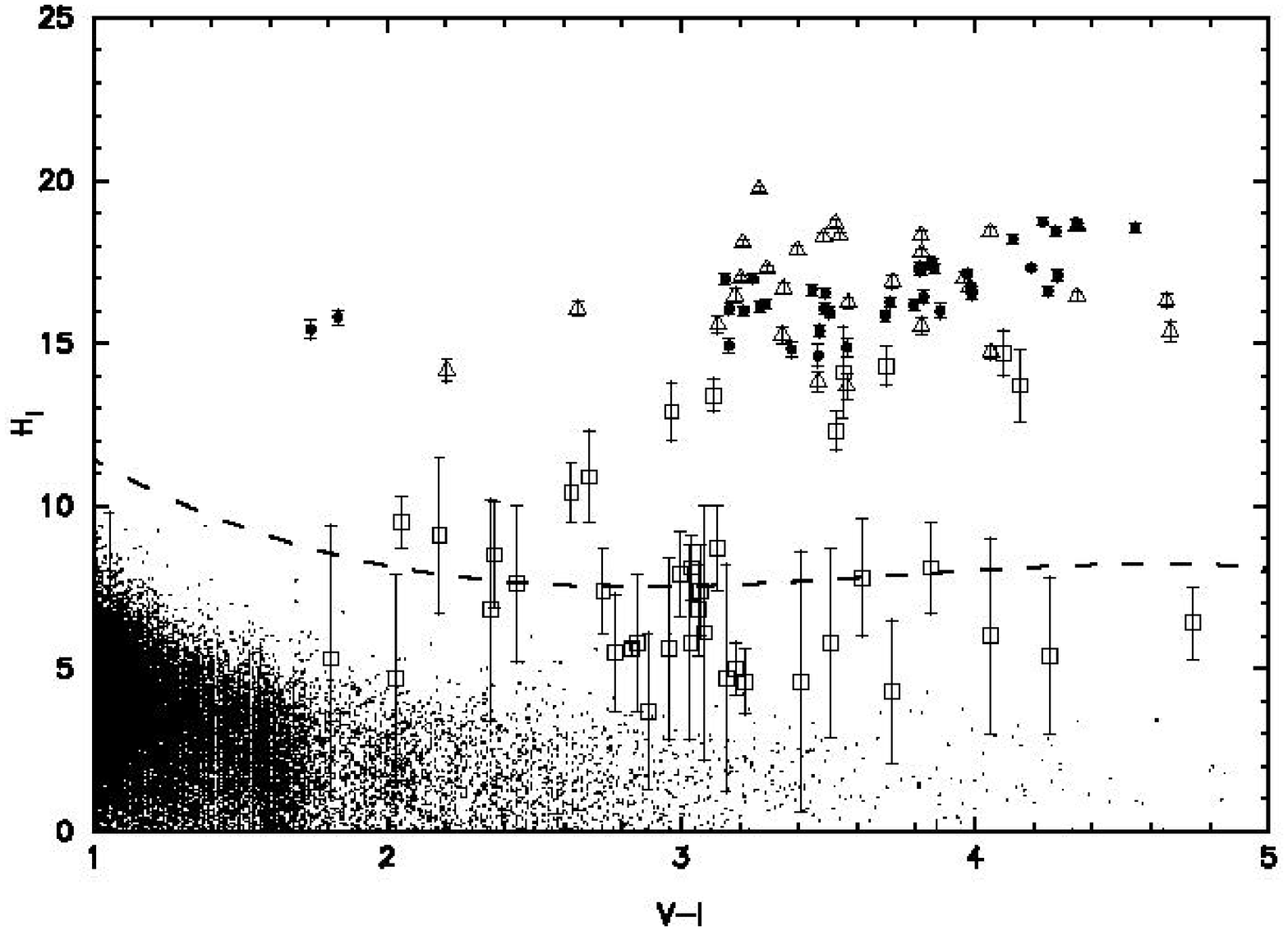,angle=-90} 
\psfig{height=6.5cm,file=ms2905f5.ps,angle=-90} 
\caption{$I$ band reduced proper motions vs. $V-I$ and $I-J$. 
Dashed curve: $H^{\rm max}_{\rm I}$ for giants. Objects above this 
curve must be dwarfs.
Solid circles: high-PM objects in this paper, 
Table~\ref{obs_highPMa}a, \ref{obs_highPMb}b;
Triangles: late-M dwarf candidates from Paper I (Table~2); Squares: 
low-PM objects, 
Tables~\ref{obs_lowPM1}c
\&~\ref{obs_lowPM2} . In the upper diagram, the many dots are HIPPARCOS 
single giants with $V-I \geq 1.0$ (28022 stars), all of which are located 
well below the dwarf/giant separation curve.}
\label{limit_giants}
\end{figure}
\begin{table*}
   \caption{{\bf{a)}} Observational data and Reduced Proper Motions for 
24 high-PM 
nearby late-M dwarf candidates selected in the 3600 square degrees}
    \label{obs_highPMa}
  $$
   \begin{tabular}{llllllllllll}
   \hline 
   \hline
   \noalign{\smallskip}
DENIS Name  & Other name &$\alpha_{\rm 2000}$ & $\delta_{\rm 2000}$& DENIS &$B$ &$R$ & $I$& $I-J$ & $J-K$ & $H_{\rm I}$ & $H^{\rm max}_{\rm I}$ \\
            &            &                    &                    & Epoch &    &    &    &       &      &          &             \\
(1)&(2)&(3)&(4)&(5)&(6)&(7)&(8)&(9)&(10)&(11)&(12)\\   
\hline
   \noalign{\smallskip}
 J0020231$-$234605      & LP 825-35 &00 20 23.17 & $-$23 46 05.7& 2000.589 & 20.4 & 17.5 & 14.65 & 2.39 & 0.98 & 17.2 &  7.8  \\  
 J0103119$-$535143      & ...       &01 03 11.98 & $-$53 51 43.6& 1999.879 & 20.1 & 17.5 & 14.54 & 2.27 & 1.19 & 16.4 &  7.7  \\  
 J0120491$-$074103      & ...       &01 20 49.15 & $-$07 41 03.5& 1999.926 & **   & 19.4 & 15.71 & 2.77 & 1.19 & 16.0 &  7.9  \\  
 J0144318$-$460432      & ...       &01 44 31.88 & $-$46 04 32.1& 1999.882 & 19.3 & 16.7 & 14.10 & 2.19 & 0.81 & 14.6 &  7.6  \\  
 J0218579$-$061749      & LP 649-93 &02 18 57.90 & $-$06 17 49.7& 2000.551 & 21.5 & 19.0 & 15.56 & 2.65 & 1.22 & 18.5 &  7.9  \\  
 J0235495$-$071121      & ...       &02 35 49.56 & $-$07 11 21.1& 1999.912 & 21.0 & 18.0 & 14.71 & 2.32 & 1.01 & 17.1 &  7.7  \\  
 J0306115$-$364753      & ...       &03 06 11.57 & $-$36 47 53.2& 1999.975 & 20.9 & 17.7 & 14.41 & 2.79 & 1.03 & 18.7 &  7.9  \\  
 J0320588$-$552015      & ...       &03 20 58.85 & $-$55 20 15.8& 1999.890 & 20.1 & 17.2 & 14.30 & 2.23 & 1.03 & 17.3 &  7.7  \\  
 J0351000$-$005244      & GJ 3252   &03 51 00.03 & $-$00 52 44.6& 1999.907 & 19.8 & 16.7 & 13.75 & 2.55 & 0.99 & 17.1 &  7.9  \\  
 J0517377$-$334903      & ...       &05 17 37.70 & $-$33 49 03.2& 1999.962 & 21.1 & 18.2 & 14.93 & 2.89 & 1.19 & 18.7 &  7.8  \\  
 J1006319$-$165326      & LP 789-23 &10 06 31.99 & $-$16 53 26.3& 2000.164 & 20.3 & 17.4 & 14.55 & 2.44 & 1.14 & 17.4 &  7.8  \\  
 J1021513$-$032309      & LP 610-5  &10 21 51.36 & $-$03 23 09.6& 2000.260 & 20.0 & 17.8 & 14.55 & 2.31 & 0.88 & 16.5 &  7.7  \\  
 J1048126$-$112009      & GJ 3622   &10 48 12.64 & $-$11 20 09.8& 2000.200 & 16.9 & 14.7 & 11.25 & 2.30 & 0.98 & 17.3 &  7.7  \\  
 J1106569$-$124402      & LP 731-47 &11 06 56.91 & $-$12 44 02.2& 2000.247 & 19.9 & 17.3 & 14.18 & 2.41 & 0.82 & 16.7 &  7.8  \\  
 J1141440$-$223215      & ...       &11 41 44.06 & $-$22 32 15.1& 2000.258 & 21.7 & 19.1 & 15.42 & 2.72 & 1.21 & 18.6 &  7.9  \\  
 J1145354$-$202105      & LP 793-34*&11 45 35.40 & $-$20 21 05.2& 2000.241 & 19.0 & 16.6 & 13.84 & 2.15 & 0.88 & 14.9 &  7.6  \\  
 J1147421$+$001506      & GJ 3686B  &11 47 42.11 & $+$00 15 06.4& 2000.197 & 18.6 & 15.7 & 13.19 & 2.06 & 0.96 & 15.4 &  7.5  \\  
 J1155429$-$222458      & LP 851-346&11 55 42.94 & $-$22 24 58.2& 1996.208 & 19.6 & 16.8 & 13.48 & 2.58 & 1.05 & 16.6 &  7.9  \\  
 J1201421$-$273746      & LP 908-5  &12 01 42.10 & $-$27 37 46.5& 1999.129 & 19.8 & 16.3 & 14.30 & 2.21 & 0.91 & 17.0 &  7.6  \\  
 J1250526$-$212113      & ...       &12 50 52.66 & $-$21 21 13.9& 2000.249 & 19.3 & 16.8 & 13.78 & 2.59 & 1.11 & 17.5 &  7.9  \\  
 J1610584$-$063132      & LP 684-33 &16 10 58.45 & $-$06 31 32.2& 2000.553 & 18.5 & 16.0 & 13.46 & 2.08 & 1.09 & 14.8 &  7.5  \\  
 J2132297$-$051158      & LP 698-2  &21 32 29.76 & $-$05 11 58.9& 2000.408 & 19.1 & 16.3 & 13.52 & 2.12 & 1.13 & 16.3 &  7.6  \\  
 J2205357$-$110428      & LP 759-25 &22 05 35.74 & $-$11 04 28.5& 1998.816 & 19.4 & 16.5 & 13.67 & 2.13 & 0.99 & 16.2 &  7.6  \\  
 J2337383$-$125027      & LP 763-3  &23 37 38.33 & $-$12 50 27.3& 1998.805 & 19.1 & 16.2 & 13.67 & 2.13 & 1.10 & 16.5 &  7.6  \\  
    \noalign{\smallskip}
    \hline 
   \end{tabular}
  $$
  \begin{list}{}{}
  \item[*] A companion to HIP 57361.
  \item[**] Too faint for the Schmidt plates.
  
Columns 1 \& 2: Object name in the DENIS data base, and other 
identification if available. 

Columns 3, 4 \& 5: DENIS Position with respect to equinox J2000 at DENIS 
epoch, and DENIS epoch.

Columns 6 \& 7: $B$ and $R$ photographic magnitudes.  

Columns 8, 9 \& 10: DENIS $I$-magnitude and colours.  

Columns 11 \& 12: $I$ band Reduced Proper Motion, and maximum R.P.M for
an M giant of the same $I-J$ colour.

  \end{list}
\end{table*}
\begin{table*}
\setcounter{table}{2}
   \caption{{\bf{b)}} Observational data and Reduced Proper Motions for 15 high-PM 
       nearby late-M dwarf candidates initially selected from the NLTT (same columns as Table~\ref{obs_highPMa}a)}
    \label{obs_highPMb}
  $$
   \begin{tabular}{llllllllllll}
   \hline 
   \hline
   \noalign{\smallskip}
DENIS Name  & Other name &$\alpha_{\rm 2000}$ & $\delta_{\rm 2000}$& DENIS &$B$ &$R$ & $I$& $I-J$ & $J-K$ & $H_{\rm I}$ & $H^{\rm max}_{\rm I}$ \\
            &            &                    &                    & Epoch &    &    &    &       &      &          &             \\
(1)&(2)&(3)&(4)&(5)&(6)&(7)&(8)&(9)&(10)&(11)&(12)\\   
\hline
   \noalign{\smallskip}
 J0002061$+$011536& LP 584-4  &00 02 06.18 & $+$01 15 36.6& 1998.570 & 20.8 & 18.0 & 14.80 & 2.53 & 1.11 & 18.2 &  7.9  \\  
 J0410480$-$125142& LP 714-37 &04 10 48.06 & $-$12 51 42.7& 2000.896 & 17.8 & 15.4 & 12.99 & 2.05 & 1.05 & 16.0 &  7.5  \\  
 J0440231$-$053009& LP 655-48 &04 40 23.17 & $-$05 30 09.1& 1996.044 & 19.0 & 15.8 & 13.35 & 2.61 & 1.19 & 15.9 &  7.9  \\  
 J0520293$-$231848& LP 836-41 &05 20 29.37 & $-$23 18 48.4& 1999.847 & 19.2 & 16.6 & 14.02 & 2.27 & 1.12 & 16.6 &  7.7  \\  
 J0931223$-$171742& LP 788-1  &09 31 22.30 & $-$17 17 42.4& 2000.186 & 19.2 & 16.0 & 13.36 & 2.32 & 1.01 & 15.9 &  7.7  \\  
 J1346460$-$314925& LP 911-56 &13 46 46.07 & $-$31 49 25.8& 1999.301 & 18.0 & 15.8 & 13.27 & 2.24 & 1.07 & 16.1 &  7.7  \\  
 J1504161$-$235556& LP 859-1  &15 04 16.15 & $-$23 55 56.4& 2001.436 & 20.0 & 17.8 & 14.72 & 2.70 & 1.10 & 17.3 &  7.9  \\  
 J1546115$-$251405& LP 860-30 &15 46 11.53 & $-$25 14 05.9& 2001.400 & 19.0 & 16.5 & 14.09 & 2.09 & 0.89 & 17.0 &  7.5  \\  
 J1552446$-$262313& LP 860-41 &15 52 44.61 & $-$26 23 13.7& 1999.534 & 17.7 & 15.0 & 12.61 & 2.24 & 1.07 & 16.2 &  7.7  \\  
 J1553571$-$231152& LP 860-46 &15 53 57.14 & $-$23 11 52.2& 1996.301 & 18.4 & 16.0 & 13.64 & 2.05 & 1.02 & 16.0 &  7.5  \\  
 J1625503$-$240008& LP 862-26 &16 25 50.33 & $-$24 00 08.5& 1999.463 & 18.0 & 15.2 & 14.39 & 2.38 & 1.39 & 15.4 &  7.8  \\  
 J1641430$-$235948& LP 862-111&16 41 43.00 & $-$23 59 48.5& 2000.545 & 17.9 & 15.0 & 14.13 & 2.15 & 1.20 & 15.8 &  7.6  \\  
 J1645282$-$011228& LP 626-2  &16 45 28.20 & $-$01 12 28.8& 2000.474 & 20.1 & 17.2 & 14.28 & 2.14 & 0.94 & 16.0 &  7.6  \\  
 J1917045$-$301920& LP 924-17 &19 17 04.51 & $-$30 19 20.1& 1999.353 & 19.1 & 16.4 & 13.81 & 2.11 & 0.95 & 16.1 &  7.6  \\  
 J2151270$-$012713& LP 638-50 &21 51 27.02 & $-$01 27 13.7& 2000.718 & 17.9 & 15.6 & 13.21 & 2.02 & 0.81 & 14.9 &  7.5  \\  
    \noalign{\smallskip}
    \hline 
   \end{tabular}
  $$
  \begin{list}{}{}
  \item[] 
  \end{list}
\end{table*}
\begin{table*}
\setcounter{table}{2}
   \caption{{\bf{c)}} Observational information and Reduced Proper Motions for the 11 
low-PM ($\mu < 0.1\arcsec$yr$^{\rm -1}$) red dwarfs candidates
   with $H_{\rm I} - H_{\rm I}^{\rm max} > 1\sigma$ in the full 5700 square degrees}
    \label{obs_lowPM1}
  $$
   \begin{tabular}{lllllllllllll}
   \hline 
   \hline
   \noalign{\smallskip}
DENIS Name  &$\alpha_{\rm 2000}$ & $\delta_{\rm 2000}$& DENIS & $B$ &  $R$ & $I$ & $I-J$ & $J-K$& $H_{\rm I}$ & err    &$H^{\rm max}_{\rm I}$& Ref  \\
            &                    &                    & Epoch &     &      &     &       &      &             & $H_{\rm I}$&                 &      \\
  (1)&(2)&(3)&(4)&(5)&(6)&(7)&(8)&(9)&(10)&(11)&(12)&(13)\\ 
 \hline
   \noalign{\smallskip}
 J0013093$-$002551 & 00 13 09.34 & $-$00 25 51.5  & 1999.838 & 19.8 & 17.2 & 14.37 & 2.22 & 0.88 & 14.3 &  0.6 &  7.6 & a  \\
 J0100021$-$615627 & 01 00 02.13 & $-$61 56 27.1  & 1999.964 & 21.8 & 17.8 & 15.01 & 2.42 & 0.94 & 14.7 &  0.7 &  7.8 & a  \\
 J0436278$-$411446 & 04 36 27.84 & $-$41 14 46.9  & 1999.893 & **   & **   & 16.04 & 2.92 & 1.12 & 12.8 &  2.1 &  7.8 & a  \\
 J0518113$-$310153 & 05 18 11.32 & $-$31 01 53.0  & 2000.011 & 19.5 & 16.8 & 14.17 & 2.30 & 1.00 & 12.3 &  0.6 &  7.7 & a  \\
 J1236396$-$172216 & 12 36 39.61 & $-$17 22 16.9  & 1999.384 & 18.2 & 16.2 & 13.91 & 2.14 & 1.14 & 12.9 &  0.9 &  7.6 & a  \\
 J1538317$-$103850 & 15 38 31.70 & $-$10 38 50.6  & 2000.414 & 18.3 & 16.4 & 14.36 & 2.18 & 0.95 & 10.9 &  1.4 &  7.6 & a  \\
 J1552237$-$033520 & 15 52 23.78 & $-$03 35 20.7  & 1999.534 & 15.8 & 13.2 & 12.02 & 2.07 & 1.37 &~\,9.5&  0.8 &  7.5 & a  \\
 J1553186$-$025919 & 15 53 18.65 & $-$02 59 19.3  & 1999.581 & 17.0 & 15.1 & 13.12 & 2.04 & 1.36 & 10.4 &  0.9 &  7.5 & a  \\
 J2022480$-$564556 & 20 22 48.01 & $-$56 45 56.8  & 2000.477 & 19.2 & 15.8 & 13.81 & 2.06 & 0.80 & 13.4 &  0.5 &  7.5 & a  \\
 J2206227$-$204706*& 22 06 22.78 & $-$20 47 06.0  & 1999.611 & 20.1 & 17.9 & 15.09 & 2.67 & 1.22 & 14.1 &  1.4 &  7.9 & a  \\
 J2226443$-$750342 & 22 26 44.36 & $-$75 03 42.7  & 1999.814 & 21.5 & 18.3 & 15.20 & 2.84 & 1.20 & 13.7 &  1.1 &  7.9 & a  \\
    \noalign{\smallskip}
    \hline 
   \end{tabular}
  $$
  \begin{list}{}{}
  \item[*] Previously listed by Gizis et al. (\cite{gizis})
  \item[**] Too faint for the plate.
  
 Columns 1, 2, 3 \& 4 : DENIS name, position with respect to equinox J2000 
            at DENIS epoch, and DENIS epoch.

 Columns 5 \& 6: $B$ and $R$ photographic magnitudes.

 Columns 7, 8 \& 9: DENIS $I$-magnitude and colours.  

 Columns 10 \& 11 : $H_{\rm I}$ $I$-band reduced proper, and its standard error.   

 Column 12: Maximum R.P.M. for a giant of the same $I-J$ colour.

 Column 13: References for the proper motion and the $B$ and $R$ photometry: 
(a) our measurements; 
(b) $B$ and $R$ from the USNO-A2.0 catalogue
  (Monet at al. \cite{monet}) and proper motion from the UCAC1 catalogue 
  (Zacharias et al. \cite{zacharias});
(c) $B$ and $R$ from the USNO-A2.0 catalogue (Monet at al. \cite{monet}) and 
  proper motion from the Tycho-2 catalogue (H\o g et al. \cite{hog}).
  \end{list}
\end{table*}
%

\section{Reduced proper motions}
In Paper I probable giants were rejected on a proper motion cutoff, by
requiring $\mu \geq 0.1\arcsec$yr$^{\rm -1}$. This criterion, while effective,
is not optimal, in that it completely ignores the photometric
information: an apparently fainter star, everything else being equal,
is farther away than a brighter one, and is thus on average expected
to have a smaller proper motion. The combination of kinematic and 
photometric information embodying that simple idea is the Reduced 
Proper Motion (RPM), extensively used by Luyten and initially coined
by Hertzsprung. The RPM is defined in terms of the observable parameters as:
\begin{eqnarray}
   H & = & m + 5 + 5 \log \mu \label{eq2}
\end{eqnarray}
where $m$ is the apparent magnitude in a given photometric band and 
$\mu$ is the total proper motion in arc-sec~yr$^{\rm -1}$. Its usefulness
becomes more apparent after it is rephrased in terms of 
intrinsic stellar parameters, to:
\begin{eqnarray}
   H & = &  M + 5 \log (V_{\rm t}/4.74)\label{eq3}
\end{eqnarray}
where $M$ is the absolute magnitude in the same photometric band and 
$V_{\rm t}$ is the tangential velocity (km/s). Under this form it is 
clear that, unless it serendipitously has a very unusually 
low tangential velocity, a dwarf will have a much larger RPM than any giant.
Subdwarfs have even larger RPMs than normal dwarfs, through a combination of
fainter magnitudes (at a given colour) and a larger velocity dispersion.
RPM vs colour plots are therefore extremely effective at statistically 
separating giants, dwarfs, subdwarfs, and white dwarfs. 

The largest possible $V_{\rm t}$ for a star bound to the Galaxy is that
of a retrograde star orbiting at the escape velocity, and located 
in the direction of either the galactic center or anticenter:
\begin{eqnarray}
  V_{\rm max} & = & V_{\rm e} + V_{\rm LSR} + V_{\rm \odot}  \label{eq4}
\end{eqnarray}
where: 
\[
 \begin{array}{lp{0.8\linewidth}}
  V_{\rm e}     & $\sim$500$\pm$40 km/s is the escape velocity  
   in the solar neighbourhood (Leonard \& Tremaine \cite{leonard}, 
   Meillon \cite{meillon}) \\
  V_{\rm LSR}   & $\sim$220~km/s (Kerr \& Lynden-Bell \cite{kerr}) is the
   rotation velocity of the Local Standard of Rest\\
  V_{\rm \odot} & $\sim$5~km/s (Dehnen \& Binney \cite{dehnen}) is the solar
   velocity relative to the LSR. 
 \end{array}
\]
Taking safe margins on all components, $V_{\rm max}$ is thus at most 
$800~km/s$. For a given stellar luminosity, and in the $I$ photometric
band,  this translates into a maximum RPM of:
\begin{eqnarray}
   H_{\rm I}^{\rm max} & = & M_{\rm V} - (V-I) + 5 \log (V_{\rm max}/4.74)  
    \label{eq5} 
\end{eqnarray}
To estimate the maximum RPM for giants at a given colour, we fitted 
polynomial functions to the $V-I$ and $I-J$ colours data from Th\'e et al. 
(\cite{the}), Bessell \& Brett (\cite{bessell88}), and used Equation 
(\ref{eq5}) to obtain:
\begin{eqnarray}
   H_{\rm I}^{\rm max} & = & 18.82-9.97(V-I)+2.83(V-I)^{2} \nonumber \\
                 &   &-~0.25(V-I)^{3} \label{eq6}                
\end{eqnarray}
for $V-I$ $\in$ [1.5,5.0]
\begin{eqnarray}
   H_{\rm I}^{\rm max} & = & 18.97-17.04(I-J)+8.08(I-J)^{2} \nonumber \\
                 &   & -~1.22(I-J)^{3} \label{eq7}                 
\end{eqnarray}
for $I-J$ $\in$ [1.0,3.0]

Fig.~\ref{limit_giants} shows the resulting $H_{\rm I}^{\rm max}$ vs. 
($V-I$) and  vs. ($I-J$) curves, and our candidates. 

To produce the ($H_{\rm I}$, $V-I$) diagram, we interpolated a very
approximate $V$-magnitude from the the $B$ and $R$ photographic magnitudes. 
This is obviously very crude, but nonetheless proves adequate:
the $H_{\rm I}^{\rm max}$ vs $V-I$ curve for giants is fairly flat, so 
that even large errors on $V-I$ do not significantly affect the position
relative to the curve. As an illustration of the very effective
giant/dwarf separation in RPM plots, the diagram also displays 
28000 single Hipparcos giants with adequate colour information, which all do 
lay well below the giants curve.

\begin{figure*}
\resizebox{16.0cm}{!}{\includegraphics{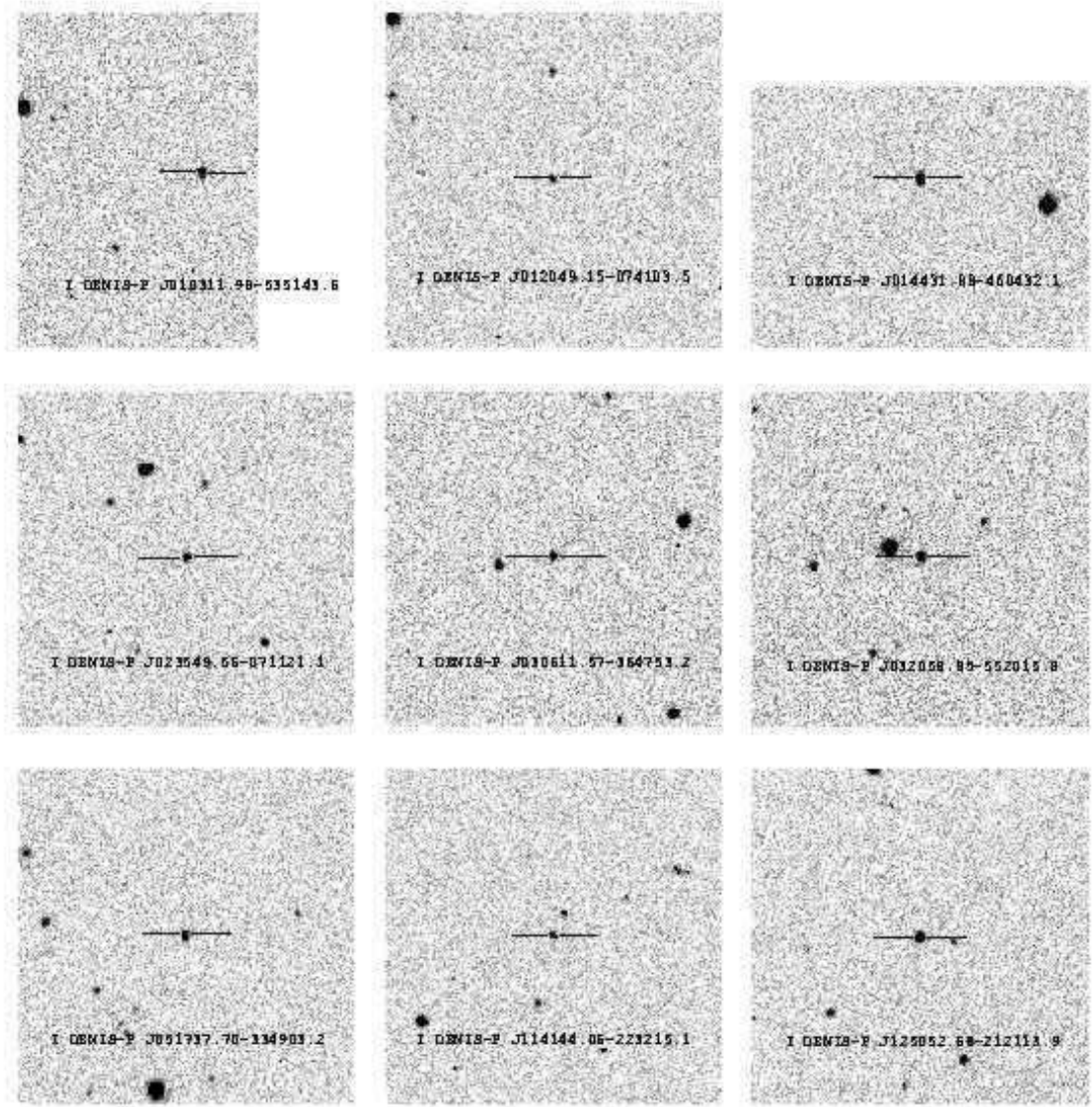}}
\caption{$I$-band finding charts for the 9 new high-PM objects listed in 
Table~\ref{pma}a. The charts are $\sim~4.0'~\times~4.0'$, with North up 
and East to the left. No finding charts are provided for the lower proper
motion objects, which are easily identified from their accurate coordinates.}
\label{charts}
\end{figure*}

We divide the photometric candidates into 3 categories, plotted
in Fig.~\ref{limit_giants}, according to their
position relative to the $H_{\rm I}^{\rm max}$ curve:
\begin{itemize}
\item[\bf\Huge{.}]  Stars with $\mu > 0.1\arcsec$yr$^{-1}$ 
are listed in Tables~\ref{pma}a and \ref{obs_highPMa}a (24 objects). As 
expected from the conservative limits used in Paper I, they are well 
above the giants 
curve, and have standard errors on $H_{\rm I}$ of $\sim 0.1$.
Fig.~\ref{charts} gives finding charts for the 9 completely new objects.
\item[\bf\Huge{.}]  Stars with $\mu < 0.1\arcsec$yr$^{-1}$, but with 
$H_{\rm I} - H_{\rm I}^{\rm max} > 1\sigma$, with $\sigma$
the uncertainty on $H_{\rm I}$ (11 objects, Table~\ref{obs_lowPM1}c). 
Their proper motions (Table~\ref{low_pm}c) have large relative
uncertainties, and, because of the logarithmic $\mu$ dependency,
their RPM measurements are thus quite noisy. A few of the noisiest and/or
closest to the giants curve might possibly be giants, but the vast majority
are dwarfs. One, DENIS-P~J2206227$-$204706, detected in 
Paper~I and ignored there because of its small proper motion,  
was in fact independently recognized as a late-M dwarf by Gizis et al. 
(\cite{gizis}). 
\item[\bf\Huge{.}] Stars with $H_{\rm I} - H_{\rm I}^{\rm max} < 1\sigma$ 
(Table~\ref{obs_lowPM2}),
are overwhelmingly giants, with a minor admixture of very low tangential
velocity dwarfs. The well known K5 dwarf Gl~710, for instance, lies outside 
our spatial and colour coverage, but with $H_{\rm I}$ = 0.23 it otherwise lays
firmly within the ``giants'' region of the RPM diagram. This list (52 objects)
includes a number of bright stars referenced as 
giants in the SIMBAD database. Our measured proper motions for those stars 
are usually not significant. This results in error bars on $H_{\rm I}$ that
are occasionally so large (up to 5~magnitudes) that some objects could
not be included in Fig.~\ref{limit_giants} without obliterating the diagram. 
Whenever possible (i.e. for the brightest
objects), we therefore replaced our own measurements by the much better 
proper motions available in the UCAC1 (Zacharias et al. \cite{zacharias}) 
and Tycho-2 (H\o g et al. \cite{hog}) catalogues.
\end{itemize}

The 114 star sample identified by the photometric criteria
($2 \leq I-J \leq 3$, $d_{\rm phot} < 30$~pc) within the 5700~square 
degrees search area can therefore be divided into: 
%
\begin{itemize}
 \item[\bf\Huge{.}] 50 new nearby late-M dwarfs, consisting of:
   \begin{itemize}
    \item[+] 18 stars already found in high-PM catalogues (NLTT, WT,...) but 
       without previous distance estimate.
    \item[+] 32 completely new discoveries (13 in Paper I; 19 in this paper).
   \end{itemize}
 \item[\bf\Huge{.}] 12 previously known nearby stars (2 in Paper I; 10
      in this one).
 \item[\bf\Huge{.}] 52 probable giants, or dwarfs with very small PM.
\end{itemize}
The 50 new nearby late-M dwarfs represent a very significant addition 
to the known sample of 12 in this part of the sky. Our setting of the 
limiting distance to 30~pc rather than 25~pc (to avoid losing true 
$d<25$~pc to distance errors) accounts for some but not most of this increase. 

The 15 high-PM red dwarfs initially selected from NLTT are also listed in 
Tables~\ref{pmb}b \&~\ref{obs_highPMb}b.

Tables~\ref{disa}a and \ref{disb}b summarize the available physical 
parameters of the red dwarfs candidates listed in Table~\ref{obs_highPMa}a 
and \ref{obs_highPMb}b (DENIS origin) and \ref{obs_lowPM1}c (NLTT origin):
absolute magnitude $M_{\rm I}$, distance, tangential velocity, and the 
approximate effective  temperature derived from $I-J$ (Section~7). 
Two new late-M dwarfs have distance estimates
within 10~pc in this paper: DENIS-PJ1552237\-$-$033520 and LP 860-41 
(DENIS-P J1552446\-$-$262313). 
Five additional new stars are closer than 15~pc: DENIS-P J0306115\-$-$364753; 
LP 851-346 (DENIS-P J1155429\-$-$222458); DENIS-P J1250526\-$-$212113; LP 788-1 
(DENIS-P J0931223\-$-$171742) 
and LP 911-56 (DENIS-P J1346460\-$-$314925).

Table~\ref{comparaison} compares our distance determinations with 
litterature values for the 10~stars with a previous measurement or 
estimate. The agreement is generally good, except for a slight systematic 
discrepancy with Cruz \& Reid \cite{cruz}: for the 5 stars in common 
the Cruz \& Reid distances are significantly larger. For 
the one star with three determinations, LP 655-48, our estimate and that of 
McCaughrean et al. \cite{mccaughrean} agree and are both smaller than
the Cruz \& Reid distance.
%

%
 \begin{table*}
\setcounter{table}{3}
   \caption{Observational informations and Reduced Proper Motions for the 
     52 probable giants with $H_{\rm I} - H_{\rm I}^{\rm max} < 1\sigma$ 
in the full 5700 square degrees
     (same columns as Table~\ref{obs_lowPM1}c)}
    \label{obs_lowPM2}
  $$
   \begin{tabular}{lllllllllllll}
   \hline 
   \hline
   \noalign{\smallskip}
DENIS Name  &$\alpha_{\rm 2000}$ & $\delta_{\rm 2000}$& DENIS & $B$ &  $R$ & $I$ & $I-J$ & $J-K$& $H_{I}$ & err        & $H^{\rm max}_{\rm I}$& Ref \\
            &                    &                    & Epoch &     &      &     &       &      &         & $H_{\rm I}$ &                 &      \\
   \hline
   \noalign{\smallskip}
 J0103401$-$854203  & 01 03 40.19 & $-$85 42 03.7  & 1996.978 & 11.4 &~\,9.0&~\,9.25& 2.19 & 0.92 &~\,6.4 &~\,0.3 &  7.6 & b   \\
 J0134067$-$101403$^{\rm g}$ & 01 34 06.71 & $-$10 14 03.6  & 2000.660 & 13.3 & 10.9 & 11.38 & 2.01 & 1.14 &~\,4.7 &~\,2.3 &  7.5 & b   \\
 J0136144$-$082710  & 01 36 14.44 & $-$08 27 10.5  & 1999.940 & 18.4 & 15.1 & 14.01 & 2.47 & 1.16 &~\,9.1 &~\,2.4 &  7.8 & a   \\
 J0250072$-$860930$^{\rm g}$ & 02 50 07.20 & $-$86 09 30.0  & 1999.712 & 13.5 & 11.4 &~\,9.26& 2.09 & 1.41 &~\,5.8 &~\,2.1 &  7.5 & b   \\
 J0441247$-$271453  & 04 41 24.70 & $-$27 14 53.6  & 1999.063 & 11.3 &~\,9.3&~\,8.92& 2.19 & 1.20 &~\,4.8 &~\,0.3 &  7.6 & b   \\
 J0451504$-$750335$^{\rm g}$ & 04 51 50.48 & $-$75 03 35.7  & 1998.816 & 15.0 & 13.8 & 13.69 & 2.14 & 2.42 &  12.1 &~\,1.1 &  7.6 & a   \\
 J0457108$-$131240  & 04 57 10.85 & $-$13 12 40.3  & 1996.060 & 14.1 & 13.2 & 10.57 & 2.16 & 1.26 &~\,5.6 &~\,2.8 &  7.6 & a   \\
 J0504267$-$744821  & 05 04 26.74 & $-$74 48 21.8  & 1996.964 & 19.3 & 16.0 & 13.79 & 2.11 & 1.36 &~\,5.5 & 32.6  &  7.6 & a   \\
 J0538515$-$645534$^{\rm g}$ & 05 38 51.59 & $-$64 55 34.4  & 1996.964 & 17.3 & 14.9 & 13.59 & 2.03 & 1.51 &~\,7.7 & 17.0  &  7.5 & a   \\
 J0543339$-$782122  & 05 43 33.95 & $-$78 21 22.4  & 1996.964 & 15.5 & 13.5 & 10.67 & 2.46 & 1.44 &~\,5.8 &~\,2.9 &  7.8 & b   \\
 J0953338$-$014950  & 09 53 33.87 & $-$01 49 50.2  & 2000.164 & 16.3 & 13.7 & 10.31 & 2.25 & 1.37 &~\,5.4 &~\,2.4 &  7.7 & a   \\
 J1021323$-$204407  & 10 21 32.30 & $-$20 44 07.4  & 2000.263 & **   & **   & 16.09 & 2.98 & 1.11 &~\,... &~\,... &  ... &     \\
 J1034458$-$175302  & 10 34 45.89 & $-$17 53 02.5  & 2000.197 & 16.8 & 14.3 & 12.01 & 2.64 & 1.26 &~\,8.7 &~\,1.3 &  7.9 & b   \\
 J1125068$+$001513  & 11 25 06.87 & $+$00 15 13.9  & 2000.268 & 15.4 & 13.3 & 10.60 & 2.32 & 1.24 &~\,4.6 &~\,4.0 &  7.7 & a   \\
 J1221525$-$135310$^{\rm g}$ & 12 21 52.50 & $-$13 53 10.3  & 1999.148 & 17.5 & 10.7 & 10.38 & 2.56 & 1.25 &~\,0.4 & 19.6  &  7.9 & a   \\
 J1338300$-$294135  & 13 38 30.05 & $-$29 41 35.2  & 2000.129 & 16.0 & 13.9 & 11.53 & 2.35 & 1.27 &~\,6.1 &~\,3.9 &  7.7 & b   \\
 J1351326$-$291851  & 13 51 32.68 & $-$29 18 51.9  & 2000.362 & 16.4 & 13.2 & 11.19 & 2.28 & 1.30 &~\,7.4 &~\,1.4 &  7.7 & b   \\
 J1400335$-$271656  & 14 00 33.51 & $-$27 16 56.2  & 1999.348 & 14.6 & 11.5 &~\,9.69& 2.09 & 1.26 &~\,5.6 &~\,0.2 &  7.5 & a   \\
 J1405376$-$221515  & 14 05 37.64 & $-$22 15 15.0  & 1999.285 & **   & **   &~\,9.49& 2.09 & 1.29 &~\,... &~\,... &  ... &     \\
 J1409294$-$164227  & 14 09 29.49 & $-$16 42 27.0  & 2000.510 & 15.5 & 13.5 & 10.46 & 2.24 & 1.29 &~\,4.3 &~\,2.2 &  7.7 & b   \\
 J1427297$-$264040  & 14 27 29.71 & $-$26 40 40.8  & 1999.419 & 15.9 & 10.8 &~\,9.68& 2.12 & 1.20 &~\,5.5 &~\,1.8 &  7.6 & b   \\
 J1437524$-$183824  & 14 37 52.45 & $-$18 38 24.0  & 2000.414 & 14.2 & 12.0 & 12.86 & 2.10 & 0.92 &~\,8.3 &~\,2.1 &  7.5 & b   \\
 J1503320$-$113217$^{\rm g}$ & 15 03 32.06 & $-$11 32 17.3  & 2000.551 & 18.0 & 15.0 & 11.25 & 2.81 & 1.38 &~\,6.4 &~\,1.1 &  7.9 & b   \\
 J1503339$-$185239  & 15 03 33.92 & $-$18 52 39.1  & 2000.551 & 14.5 & 12.6 & 10.39 & 2.05 & 1.26 &~\,3.8 &~\,7.0 &  7.5 & b   \\
 J1510397$-$212524  & 15 10 39.72 & $-$21 25 24.9  & 1999.384 & 14.1 & 12.6 & 10.06 & 2.22 & 1.18 &~\,6.8 &~\,1.4 &  7.6 & b   \\
 J1512333$-$103241  & 15 12 33.30 & $-$10 32 41.3  & 2000.277 & **   & **   & 16.00 & 2.90 & 1.18 &~\,... &~\,... &  ... &     \\
 J1525014$-$032359  & 15 25 01.46 & $-$03 23 59.5  & 1999.351 & 14.1 & 11.6 &~\,9.25& 2.09 & 1.08 &~\,5.0 &~\,0.8 &  7.5 & c   \\
 J1539153$+$004404$^{\rm g}$ & 15 39 15.30 & $+$00 44 04.0  & 2000.411 & 15.8 & 12.5 & 11.78 & 2.10 & 1.31 &~\,5.3 &~\,4.1 &  7.5 & a   \\
 J1552551$-$045215  & 15 52 55.19 & $-$04 52 15.3  & 1999.534 & 14.1 & 11.8 & 10.21 & 2.01 & 1.38 &~\,8.5 &~\,1.6 &  7.5 & a   \\
 J1601227$-$093816  & 16 01 22.79 & $-$09 38 16.2  & 2000.323 & 14.3 & 11.6 & 10.47 & 2.07 & 1.25 &~\,4.7 &~\,3.2 &  7.5 & a   \\
 J1615446$-$040526  & 16 15 44.69 & $-$04 05 26.2  & 1999.353 & 14.3 & 11.5 &~\,9.67& 2.03 & 1.19 &~\,4.2 &~\,7.3 &  7.5 & a   \\
 J1952020$-$553558  & 19 52 02.08 & $-$55 35 58.8  & 2000.477 & 16.6 & 13.6 & 11.26 & 2.35 & 1.27 &~\,5.1 &~\,6.2 &  7.7 & b   \\
 J2004401$-$395151  & 20 04 40.14 & $-$39 51 51.7  & 2000.515 & 14.4 & 12.1 &~\,9.72& 2.00 & 1.31 &~\,4.7 &~\,3.5 &  7.5 & b   \\
 J2015585$-$712313  & 20 15 58.52 & $-$71 23 13.2  & 2000.529 & 17.2 & 11.1 & 10.89 & 2.59 & 1.26 &~\,4.5 &~\,6.3 &  7.9 & b   \\
 J2016341$-$772709  & 20 16 34.12 & $-$77 27 09.4  & 2000.537 & 17.3 & 14.5 & 11.58 & 2.44 & 1.29 &~\,8.1 &~\,1.4 &  7.8 & a   \\
 J2023115$-$283921  & 20 23 11.54 & $-$28 39 21.5  & 2000.477 & 14.7 & 12.5 & 10.21 & 2.19 & 1.28 &~\,5.8 &~\,3.0 &  7.6 & b   \\
 J2024329$-$294402$^{\rm g}$ & 20 24 32.96 & $-$29 44 02.6  & 1999.392 & 14.8 & 14.3 & 10.45 & 2.13 & 1.26 &~\,6.0 &~\,3.0 &  7.6 & b   \\
 J2032270$-$273058  & 20 32 27.03 & $-$27 30 58.4  & 1999.534 & 15.0 & 12.3 & 10.76 & 2.45 & 1.18 &~\,7.6 &~\,2.4 &  7.8 & a   \\
 J2036432$-$170727  & 20 36 43.24 & $-$17 07 27.1  & 2000.592 & 14.9 & 11.5 &~\,9.40& 2.00 & 1.24 &~\,4.6 &~\,1.0 &  7.5 & b   \\
 J2044066$-$173457  & 20 44 06.68 & $-$17 34 57.3  & 1999.606 & 16.1 & 13.3 & 11.28 & 2.42 & 1.29 &~\,5.5 &~\,5.2 &  7.8 & b   \\
 J2055240$-$322600  & 20 55 24.07 & $-$32 26 00.8  & 1999.669 & 14.4 & 13.4 & 10.73 & 2.10 & 1.30 &~\,8.1 &~\,1.0 &  7.5 & b   \\
 J2056329$-$782540  & 20 56 32.90 & $-$78 25 40.1  & 1999.660 & 15.5 & 12.4 & 10.43 & 2.08 & 1.20 &~\,7.9 &~\,1.3 &  7.5 & b   \\
 J2058075$-$730350  & 20 58 07.55 & $-$73 03 50.4  & 1999.660 & 17.6 & 14.1 & 11.89 & 2.35 & 1.29 &~\,6.5 &~\,6.6 &  7.7 & a   \\
 J2103375$-$783831  & 21 03 37.56 & $-$78 38 31.5  & 1999.658 & 16.2 & 13.9 & 11.42 & 2.08 & 1.30 &~\,4.8 &~\,8.1 &  7.5 & b   \\
 J2107070$-$361729  & 21 07 07.01 & $-$36 17 29.8  & 1996.422 & 15.9 & 13.0 & 11.61 & 2.09 & 1.30 &~\,6.8 &~\,3.4 &  7.5 & b   \\
 J2108330$-$212051$^{\rm g}$ & 21 08 33.06 & $-$21 20 51.3  & 2000.567 & 17.0 & 11.7 &~\,9.80& 2.13 & 1.27 &~\,7.8 &~\,1.8 &  7.6 & a   \\
 J2124575$-$341655  & 21 24 57.51 & $-$34 16 55.9  & 1999.559 & 18.0 & 13.7 & 13.60 & 2.37 & 1.32 &~\,8.6 &~\,6.4 &  7.8 & a   \\
 J2130021$-$815158  & 21 30 02.15 & $-$81 51 58.6  & 1999.510 & 15.2 & 12.2 & 10.33 & 2.17 & 1.37 &~\,3.4 &~\,9.6 &  7.6 & b   \\
 J2141290$-$844040  & 21 41 29.02 & $-$84 40 40.1  & 2000.616 & 14.9 & 12.0 & 11.02 & 2.11 & 1.29 &~\,5.4 &~\,5.3 &  7.6 & b   \\
 J2203522$-$593300  & 22 03 52.29 & $-$59 33 00.7  & 1999.649 & 12.5 & 12.2 & 11.29 & 2.43 & 1.16 &~\,7.8 &~\,2.0 &  7.8 & b   \\
 J2225004$-$121606  & 22 25 00.48 & $-$12 16 06.9  & 1999.447 & 15.0 & 12.4 & 10.38 & 2.25 & 1.19 &~\,3.7 &~\,2.4 &  7.7 & b   \\
 J2239371$-$715950  & 22 39 37.13 & $-$71 59 50.0  & 2000.616 & 14.5 & 12.1 & 10.17 & 2.03 & 1.27 &~\,7.4 &~\,1.3 &  7.5 & b   \\
    \noalign{\smallskip}
    \hline 
   \end{tabular}
  $$
  \begin{list}{}{}
  \item[**] Too faint for the plate.
  \item[$^{\rm g}$] Previously known giants.
  \end{list}
\end{table*}
\begin{table*}
   \caption{{\bf{a)}} Estimated distances and other parameters for the 24 high-PM
   of Tables~\ref{pma}a \&~\ref{obs_highPMa}a and 11 low-PM DENIS red 
dwarf candidates of Tables~\ref{low_pm}c \&~\ref{obs_lowPM1}c.}
    \label{disa}
  $$
   \begin{tabular}{llllll|llllll}
   \hline
   \hline
   \noalign{\smallskip}
DENIS objects     & $M_{\rm I}$& D     & $V_{\rm t}$ & $T_{\rm eff}$ & Ref &\
DENIS objects     & $M_{\rm I}$& D     & $V_{\rm t}$ & $T_{\rm eff}$ & Ref  \\
                  &            &   [pc]&     [km/s]  &   [K]     &     & \
                  &            &   [pc]&     [km/s]  &   [K]     &       \\ 
   \hline
   \noalign{\smallskip}
J0013093$-$002551* & 12.59 & 22.7 & 10.4 & 2630 &     & J1141440$-$223215  & 13.56 & 23.6 & 47.4 & 2350 &     \\
J0020231$-$234605  & 12.99 & 21.4 & 33.4 & 2520 &     & J1145354$-$202105  & 12.39 & 19.5 & 14.9 & 2670 &     \\
J0100021$-$615627* & 13.05 & 24.6 & 10.3 & 2510 &     & J1147421$+$001506  & 12.10 & 16.5 & 21.5 & 2740 &  a  \\
J0103119$-$535143  & 12.72 & 23.1 & 26.1 & 2600 &     & J1155429$-$222458  & 13.34 & 10.7 & 21.3 & 2420 &     \\
J0120491$-$074103  & 13.64 & 26.0 & 14.3 & 2320 &     & J1201421$-$273746  & 12.57 & 22.2 & 36.2 & 2630 &     \\
J0144318$-$460432  & 12.51 & 20.8 & 12.5 & 2650 &     & J1236396$-$172216* & 12.36 & 20.4 &~\,6.0& 2680 &     \\
J0218579$-$061749  & 13.45 & 26.4 & 47.4 & 2380 &  d  & J1250526$-$212113  & 13.35 & 12.2 & 32.2 & 2410 &     \\
J0235495$-$071121  & 12.84 & 23.6 & 33.4 & 2570 &     & J1538317$-$103850* & 12.48 & 23.8 &~\,2.3& 2650 &     \\
J0306115$-$364753  & 13.67 & 14.0 & 48.0 & 2310 &     & J1552237$-$033520* & 12.14 &~\,9.5&~\,1.4& 2730 &     \\
J0320588$-$552015  & 12.62 & 21.7 & 40.9 & 2620 &     & J1553186$-$025919* & 12.03 & 16.5 &~\,2.2& 2750 &     \\
J0351000$-$005244  & 13.29 & 12.4 & 28.0 & 2430 &  a  & J1610584$-$063132  & 12.17 & 18.1 & 16.0 & 2720 &     \\
J0436278$-$411446* & 13.92 & 26.6 &~\,2.8& 2250 &     & J2022480$-$564556* & 12.10 & 22.0 &~\,8.8& 2740 &     \\
J0517377$-$334903  & 13.85 & 16.4 & 44.8 & 2270 &     & J2132297$-$051158  & 12.30 & 17.5 & 29.4 & 2690 &  d  \\
J0518113$-$310153* & 12.80 & 18.8 &~\,3.7& 2580 &     & J2205357$-$110428  & 12.33 & 18.5 & 27.9 & 2690 &  b  \\
J1006319$-$165326  & 13.09 & 19.6 & 34.0 & 2490 &  e  & J2206227$-$204706* & 13.48 & 21.0 &~\,6.4& 2370 &  c  \\
J1021513$-$032309  & 12.82 & 22.2 & 26.2 & 2570 &     & J2226443$-$750342* & 13.76 & 19.4 &~\,4.6& 2290 &     \\
J1048126$-$112009  & 12.80 &~\,4.9& 38.0 & 2580 &  a  & J2337383$-$125027  & 12.33 & 18.5 & 32.7 & 2690 &  d  \\
J1106569$-$124402  & 13.03 & 16.9 & 25.2 & 2510 &  b  &                    &       &      &      &      &     \\
    \noalign{\smallskip}
    \hline 
   \end{tabular}
  $$
  \begin{list}{}{}
  \item[*] Low-PM red dwarf candidates
  
  Column 1: Object name.     
  
  Columns 2, 3:  $M_{\rm I}$ absolute I-band magnitude and photometric
distance. 

  Column 4: $V_{\rm t}$ tangential velocity. The small values (all
     below 55~km/s) point to a sample dominated by disk populations.

  Column 5: $T_{\rm eff}$ effective temperature, derived from our ($I-J$, 
       $T_{\rm eff}$) calibration (see below).
  
  Column 6: Reference for a previously known trigonometric parallax: (a) 
Gliese \& Jahrei\ss ~\cite{gliesecns3} (CNS3 catalogue); for a spectrophotometric 
distance: (b) Kirkpatrick et al. \cite{kirkpatrick97}, (c) Gizis et al. \cite{gizis}, (d) Cruz \& Reid \cite{cruz},
(e) McCaughrean et al. \cite{mccaughrean};
or for a photometric distance: (f) Reid \& Cruz \cite{reid}.
  
  \end{list}
 \end{table*}
\begin{table*}
\setcounter{table}{4}
   \caption{{\bf{b)}} Estimated distances and other parameters for the 15 
DENIS red 
dwarf candidates initially selected from NLTT of 
Tables~\ref{pmb}b \&~\ref{obs_highPMb}b (same
columns as Table~\ref{disa}a).}
    \label{disb}
  $$
   \begin{tabular}{llllll}
   \hline
   \hline
   \noalign{\smallskip}
DENIS objects     & $M_{\rm I}$& D     & $V_{\rm t}$ & $T_{\rm eff}$ & Ref  \\
                  &            &   [pc]&     [km/s]  &   [K]     &       \\ 
   \hline
   \noalign{\smallskip}
J0002061$+$011536  & 13.25 & 20.4 & 46.3 & 2440 &     \\
J0410480$-$125142  & 12.07 & 15.3 & 29.0 & 2740 &  d  \\
J0440231$-$053009  & 13.39 &~\,9.8& 15.3 & 2400 &  d, e  \\
J0520293$-$231848  & 12.72 & 18.2 & 28.8 & 2600 &     \\
J0931223$-$171742  & 12.84 & 12.7 & 19.0 & 2570 &     \\
J1346460$-$314925  & 12.65 & 13.3 & 23.5 & 2620 &     \\
J1504161$-$235556  & 13.53 & 17.3 & 26.7 & 2360 &  f  \\
J1546115$-$251405  & 12.20 & 23.8 & 42.8 & 2710 &     \\
J1552446$-$262313  & 12.65 &~\,9.8& 24.4 & 2620 &     \\
J1553571$-$231152  & 12.07 & 20.6 & 29.6 & 2740 &  f  \\
J1625503$-$240008  & 12.97 & 19.2 & 14.7 & 2530 &     \\
J1641430$-$235948  & 12.39 & 22.3 & 22.8 & 2670 &     \\
J1645282$-$011228  & 12.36 & 24.2 & 25.2 & 2680 &     \\
J1917045$-$301920  & 12.27 & 20.3 & 27.0 & 2700 &     \\
J2151270$-$012713  & 11.96 & 17.8 & 18.6 & 2760 &  d  \\
    \noalign{\smallskip}
    \hline 
   \end{tabular}
  $$
  \begin{list}{}{}
  \item[] 
  \end{list}
 \end{table*}
 %
%
%
\begin{table*}
   \caption{Comparison between our photometric distances from 
     ($I-J$, $M_{\rm I}$) 
     and literature distances, based on either trigonometric parallaxes or 
     spectrophotometric distances with stated accuracies better than 
     4~pc. The Hipparcos distance quoted for LP~793-34 results from the
     parallax of its common proper motion companion, LP~793-33}
     \label{comparaison}
  $$
   \begin{tabular}{lllll}
   \hline 
   \hline
   \noalign{\smallskip}
 DENIS name            & Other Name  & Our distance & Previous distance & Source  \\
                       &             &   [pc]       &   [pc]        &         \\
   \hline
   \noalign{\smallskip}
J0351000$-$005244      &  GJ 3252    & 12.4     & 14.7~$\pm$~0.4      &  Gliese \& Jahrei\ss ~\cite{gliesecns3}   \\
J0410480$-$125142      &  LP 714-37  & 15.3     & 19.4~$\pm$~2.1      &  Cruz   \& Reid \cite{cruz}              \\
J0440231$-$053009      &  LP 655-48  & ~\,9.8   & ~\,8.0~$\pm$~1.6    &   McCaughrean et al. \cite{mccaughrean} \\
                       &             & ~\,9.8   & 15.3~$\pm$~2.6      &   Cruz   \& Reid \cite{cruz}             \\
J1048126$-$112009      &  GJ 3622    & ~\,4.9   &~\,4.5~$\pm$~0.1     &  Gliese \& Jahrei\ss ~\cite{gliesecns3}   \\
J1106569$-$124402      &  LP 731-47  & 16.9     & 18.0~$^{~+3}_{~-2}$ &  Kirkpatrick et al. \cite{kirkpatrick97} \\
J1145354$-$202105      &  LP 793-34  & 19.5     & 20.2~$\pm$~1.5      &  Hipparcos, for LP 793-33                \\
J1147421$+$001506      &  GJ 3686B   & 16.5     & 15.6~$\pm$~2.9      &  Gliese \& Jahrei\ss ~\cite{gliesecns3}   \\
J2132297$-$051158      &  LP 698-2   & 17.5     & 23.7~$\pm$~2.8      &  Cruz   \& Reid \cite{cruz}              \\
J2151270$-$012713      &  LP 638-50  & 17.8     & 21.0~$\pm$~1.5      &  Cruz   \& Reid \cite{cruz}              \\
J2337383$-$125027      &  LP 763-3   & 18.5     & 21.5~$\pm$~2.3      &  Cruz   \& Reid \cite{cruz}              \\
   \noalign{\smallskip}
    \hline 
   \end{tabular}
  $$
\end{table*}
%

\section{Sample completeness, and the local late-M dwarf density} 
Since the stars which were initially fetched from proper motion catalogues
have a very different (and poorly controlled) selection function, we restrict
the discussion in this section to the colour-selected dwarf candidates in 
the full 5700 square degrees (this paper and Paper I). We also ignore the 
52 probable giants of Table~\ref{obs_lowPM2}, which are of a different
physical nature, and therefore use a sample of 62 late-M dwarfs 
 in the density calculation:

\begin{itemize}
  \item[\bf\Huge{.}] 26 colour-selected high-PM stars from 
       the 2100 square degrees explored in Paper I ( Table~2, excluding
       4 stars fetched from the LHS outside this area).
  \item[\bf\Huge{.}] 25 colour-selected high-PM stars from
     the additional 3600 square degrees explored here 
     ( Table~\ref{obs_highPMa}a). This includes 
      the colour-selection of LHS~5165, identified in Paper~I from
      the LHS catalog and which happens to lay within 
      the additional sky coverage.
\item[\bf\Huge{.}] 11 colour-selected lower proper motion probable dwarfs, 
     found
     over the full 5700~square degrees ( Table~\ref{obs_lowPM1}c).
\end{itemize}

%
%

\begin{figure}
\psfig{height=6.5cm,file=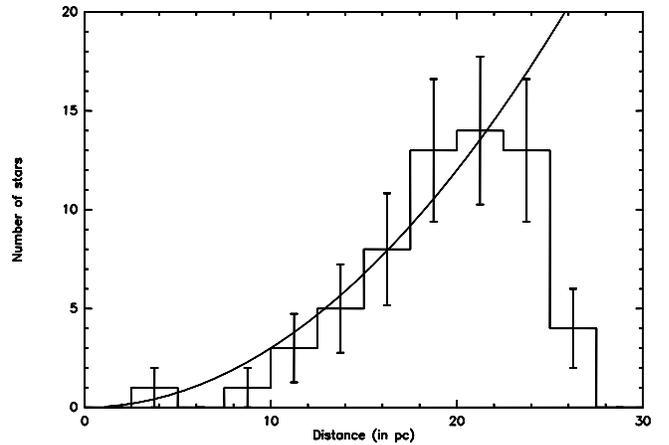,angle=-90}
\caption{Number of red dwarf candidates per 2.5~pc photometric distance bin
over 5700 square degrees. The errorbars are poissonian 1$\sigma$ errors
and the curve is the expected $d^2$ distribution, normalized at 18pc.}
\label{distribution_function}
\end{figure}

The differential photometric distance distribution of that sample 
(Fig.~\ref{distribution_function}) is well fitted by a $d^2$ distribution,
as expected for a constant-density population, out to $\sim$ 22-25~pc. The
difference from the initial 30~pc selection cutoff reflects the slightly 
different colour-magnitude relations used in the selection and in the final 
photometric distance estimate. We conservatively adopt 22~pc as the 
completeness limit of our sample, and use the 45 stars within that
distance to determine the local density of late-M dwarfs. 
 Using the Reid \& Cruz (\cite{reid}) ($I-J$,~$M_I$) relation 
would give slightly larger distances and change the completeness
distance to 25 pc. 

A sample limited by photometric distance is effectively a magnitude-limited
sample, with a colour-dependent magnitude limit. As such, and since the
colour-luminosity relation has significant dispersion, it is subject
to the well-know Malmquist bias (Malmquist \cite{malm36}), through two
separate but interrelated effects (e.g. Stobie et al. \cite{stobie89} 
and Kroupa \cite{kroupa98}):
\begin{itemize}
\item[\bf\Huge{.}] The average luminosity at a given colour is brighter for a 
magnitude-limited sample than for a volume-limited sample, since
the brighter stars are included to larger distances, hence in larger
numbers, than the fainter ones. This is the classical Malmquist bias;
\item[\bf\Huge{.}] A magnitude-limited sample includes more stars at a given
colour than the equivalent volume-limited sample for the average 
colour-luminosity relation: since the volume grows as $d^3$, the additional
volume from which brighter stars get included is larger that the missing
volume from which fainter stars are lost. 
\end{itemize}
Here we are only interested in the colour-integrated stellar density over 
$2.0 \leq I-J \leq 3.0$ (or $11.9 \leq M_{\rm I} \leq 14.0$). The first
component 
of the Malmquist bias is therefore irrelevant, since a) we do not look
for any significant luminosity resolution, and b) the luminosity function 
is sufficiently flat over the M6-M8 spectral range (Delfosse \& Forveille 
\cite{delfor00}) that a small shift in the average luminosity will not
measurably affect the resulting density. The second component of the bias, 
on the other hand, is significant. For a gaussian dispersion of the 
colour-luminosity relation it can be computed analytically 
(Stobie et al. \cite{stobie89}):
\begin{eqnarray}
\frac{{\Delta}{\Phi}}{\Phi}=\frac{1}{2}{\sigma}^2(0.6~{\ln}10)^2 \label{eq9}
\end{eqnarray}
where $\Phi$ is the luminosity function and $\sigma$ is the intrinsic rms
scatter in the colour-luminosity relation. The scatter in the $M_I$ vs 
$I-J$ relation is $\sigma \sim 0.2$ mag (Fig. \ref{fig_col_mag}), which 
corresponds to a 4\% overestimate of the stellar density. 

The mean surface density of our sample, $0.66 \pm 0.11$ objects per 100 
square degrees out to 22~pc, corresponds to an uncorrected luminosity 
function of 
$\overline{\Phi}_{\rm I}=(2.3 \pm 0.4) . 10^{-3}$ stars.$M_{\rm I}^{-1}$.pc$^{-3}$. 
After correcting
for the Malmquist bias, this becomes 
$\overline{\Phi}_{\rm I~cor}=(2.2 \pm 0.4).10^{-3}$ stars.$M_{\rm I}^{-1}$.pc$^{-3}$, averaged
over $11.9 \leq M_{\rm I} \leq 14.0$. Using relations from Leggett 
(\cite{leggett}) and Dahn et al. (\cite{dahn02}) to translate to $M_{\rm V}$,
this gives $\overline{\Phi}_{\rm V~cor}=(1.7 \pm 0.3).10^{-3}$ stars.$M_{\rm V}^{-1}$.pc$^{-3}$, 
averaged over $15.4 \leq M_{\rm V} \leq 18.7$.

\begin{figure}
\psfig{height=6.5cm,file=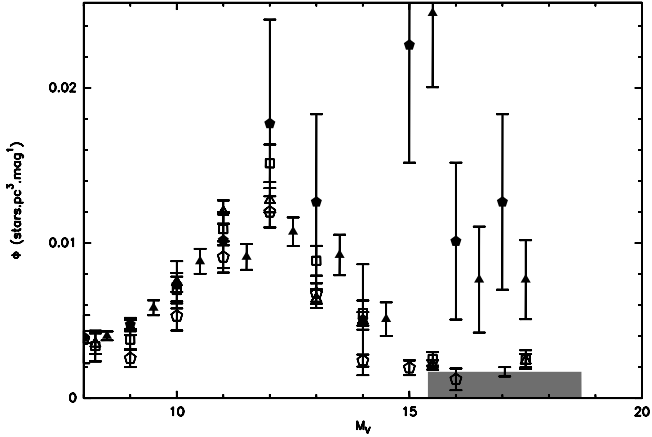,angle=-90}
\caption{The $M_V$ luminosity function. Open symbols are photometric 
luminosity function (triangles and squares are two Galactic models 
from Zheng et al. \cite{zheng01}; polygons are from Kroupa \cite{kroupa95a}).
Filled symbol represent nearby star luminosity functions (triangles from Reid, 
Gizis \& Hawley \cite{reid02}, and polygons from Kroupa \cite{kroupa95a}).
The filled grey area shows our stellar density estimate for M6 to M8 stars,
which is in excellent agreement with other photometric luminosity functions. 
}
\label{fnlum}
\end{figure}

Stellar luminosity functions for the solar neighborhood come in two kinds:
photometric luminosity functions, with somewhat uncertain distances and
luminosities estimated from colour-luminosity relations, and nearby star
luminosity functions, with distances (mostly) from trigonometric parallaxes
but with typically smaller samples and sometimes an uncertain completeness.
Fig~\ref{fnlum} compares the above stellar density measurement with
the photometric luminosity functions of Kroupa (\cite{kroupa95a}) and 
Zheng et al. (\cite{zheng01}), as well as with the nearby star luminosity 
functions of Kroupa (\cite{kroupa95a}) and Reid, Gizis \& Hawley 
(\cite{reid02}). We measure the stellar density from a nearby sample, 
as the nearby luminosity functions do, but with the distance method from
photometric luminosity functions. Our measurement is thus of 
interest to the long-standing discrepancy between the two measurement 
techniques. Our value of $\overline{\Phi}_{\rm V~cor}=(1.7 \pm
0.3).10^{-3}$stars.$M_{\rm V}^{-1}$.pc$^{-3}$ turns out to be in 
excellent agreement with all recent measurements of the photometric 
luminosity function (e.g. Fig~\ref{fnlum} and caption). The nearby
star luminosity function is, by contrast, over an order of magnitude larger. 
This clearly excludes that a local faint star overdensity can explain the 
discrepancy, as sometimes suggested in spite of serious kinematic 
difficulties.
The true explanation most likely will have to be found in a bias of
the photometric luminosity function methodology, such as the neglecting of 
unresolved binary systems (Kroupa \cite{kroupa95b}), or potentially the 
use of an incorrect colour-luminosity relation (Reid \& Gizis \cite{reid97}; 
 Delfosse \& Forveille, in preparation). 
For a constant-density population, a systematic error in the stars luminosity
function of ${\Delta}m$ results in a luminosity function 
that is incorrect by:
\begin{eqnarray}
\frac{{\Delta}\Phi}{\Phi}=0.6\ln10{\Delta}m~{\simeq}~1.38{\Delta}m
\end{eqnarray}

In the $(I-J)$ range of interest here, Fig.~\ref{fig_col_mag} shows 
that the dispersion of the calibration stars around our adopted relation 
is 0.25~magnitude. Similarly, the rms difference between our relation
and that of Reid \& Cruz (\cite{reid}) is only 0.13 magnitude, and the
maximum difference is below 0.2~magnitude. A 0.2~magnitude error on the 
color-luminosity relation is thus a conservative upper bound. This
would affect the luminosity function at the 25-30\% level at most, well
below the difference between photometric and nearby stars luminosity 
functions. We obtain a more realistic estimate of the probable star density 
error stemming from colour-luminosity uncertainties by using the Reid 
\& Cruz (\cite{reid}) calibration (Fig.~\ref{fig_col_mag}) instead of
our own. The completeness limit is then 25~pc, with 51 stars within that 
distance, for a luminosity function of 
$\overline{\Phi}_{\rm V~cor}=(1.55 \pm 0.3).10^{-3}$stars.$M_{\rm 
V}^{-1}$.pc$^{-3}$. This is just $\sim$~10\% smaller than our best estimate,
and actually well below its Poisson probable error. 

Since parallaxes out to 30~pc can be measured very 
accurately (Dahn et al. \cite{dahn02}; Henry et al. \cite{henry97}),
though certainly with significant efforts, true distances could
be measured for the present well understood sample, and would certainly
help clarifying the source of this discrepancy.


\section{Effective temperature}
\begin{figure}
\psfig{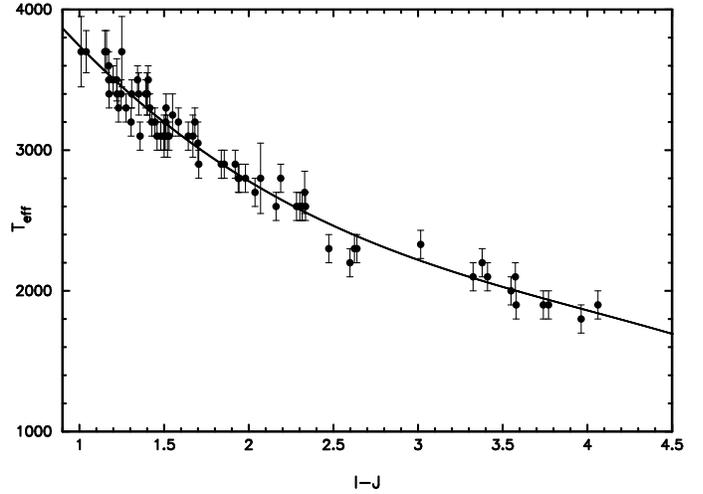}
\caption{The polynomial ($I-J$, $T_{\rm eff}$) relation, fitted to data
from Leggett, Basri et al., Tinney et al., and Bessell, see text.}
\label{Teff}
\end{figure}

The effective temperature $T_{\rm eff}$ of a star is one of its basic 
physical parameters, and we felt that a convenient rough estimate would
be useful.
We compiled
%
%
the data from Leggett et al. (\cite{leggett96}, \cite{leggett00}, 
\cite{leggett01}); Basri et al. (\cite{basri}); Tinney et al. 
(\cite{tinneya}), Bessell (\cite{bessell}), transformed when necessary to
the CIT system with the relations from Leggett (\cite{leggett}) and
Casali \& Hawarden \cite{casali}), and adjusted the following
cubic relation (Fig.~\ref{Teff}):
\begin{eqnarray}
 T_{\rm eff} & = & b_{0}+b_{1}(I-J)+b_{2}(I-J)^{2}+b_{3}(I-J)^{3} \label{eq8}
\end{eqnarray}
where $b_{0}=5297.3$, $b_{1}=-1926.3$, $b_{2}=400.0$, $b_{3}=-33
.3$  valid for $1.0 \leq I-J \leq 4.1$

It strictly speaking is only valid for CIT photometry, but 
Fig.~\ref{fig_col_mag} shows that the DENIS and CIT systems are 
sufficiently close that it provides an acceptable determination of
$T_{\rm eff}$ from DENIS photometry. 
Tables~\ref{disa}a \&~\ref{disb}b lists the effective temperatures 
derived from the DENIS $I-J$ colour index with this formula.
     
\section{Future prospects}
Compared with the cruder proper motion cutoff, selection on reduced proper
motion contributes 11 additional probable dwarfs in a sample of 62. This 
18\% fraction, which may be a lower limit if a few additional dwarfs hide
amongst the probable giants, is much larger than the 6\% loss estimated 
in Paper I. 

It is therefore important to obtain spectroscopy to make sure that all 11 
low-PM dwarf candidates are really dwarfs, 
and to determine which, if any, of the
52 probable giants are actually very low tangential velocity dwarfs.
We additionally plan to extend the systematic search to the rest of the DENIS 
data, as  they become available, as well as to the much more numerous early 
M-dwarfs candidates ($1 \leq I-J < 2$, M0-M6). A larger fraction  of those 
is probably  already known however.

\begin{acknowledgements}
We are grateful to Ren\'e Chesnel for scanning and pre-reducing the 
photographic plates. The long-term loan of POSS~I plates 
by the Leiden Observatory to Observatoire de Paris is gratefully acknowledged.
This research has made an intensive use of the Simbad and Vizier databases,
operated at CDS, Strasbourg, France.
We thank the referee for many valuable comments and suggestions.
 
\end{acknowledgements}

\end{document}